\documentclass[iop,revtex4-1]{emulateapj}
\usepackage{aas_macros}
\usepackage[dvipsnames,svgnames]{xcolor}
\usepackage{color,soul}
\usepackage{natbib}
\usepackage{hyperref}
\usepackage{amsmath}
\usepackage{xspace}
\usepackage{lineno}

\newcommand{\code}[1]{\texttt{#1}\xspace}

\newcommand{\unit}[1]{\ensuremath{\mathrm{\,#1}}\xspace}

\newcommand*\ruleline[1]{\par\noindent\raisebox{1ex}{\makebox[0.97\linewidth]{\hrulefill\quad\raisebox{-.2ex}{#1}\quad\hrulefill}}}

\newcommand{\feh}         {\mbox{[Fe/H]}}
\newcommand{\cm}{\unit{cm}}
\newcommand{\km}{\unit{km}}

\newcommand{\kpc}{\unit{kpc}}

\newcommand{\second}{\unit{s}}
\newcommand{\kms}         {\ensuremath{\km\,\second^{-1}}\xspace}

\newcommand{\msun}{\unit{M_\odot}}
\newcommand{\lsun}{\unit{L_\odot}}
\newcommand{\MeV}{\unit{MeV}}
\newcommand{\GeV}{\unit{GeV}}
\newcommand{\TeV}{\unit{TeV}}
\newcommand{\bbbar}{\ensuremath{b \bar b}\xspace}
\newcommand{\tautau}{\ensuremath{\tau^{+}\tau^{-}}\xspace}

\newcommand{\vbulk}{\ensuremath{477.2 \pm 1.2}\xspace}

\newcommand{\vgsr}{\ensuremath{235}\xspace}
\newcommand{\vdisp}{\ensuremath{3.4^{+1.2}_{-0.8} }\xspace}

\newcommand{\vbulkcariii}{\ensuremath{284.6^{+3.4}_{-3.1}}\xspace}
\newcommand{\vdispcariii}{\ensuremath{5.6^{+4.3}_{-2.1} }\xspace}
\newcommand{\vgsrb}{\ensuremath{42}\xspace}

\newcommand{\mass}{\ensuremath{1.0^{+0.8}_{-0.4} \times 10^{6}}\xspace}
\newcommand{\masstolight}{\ensuremath{369^{+309}_{-161}}\xspace}
\newcommand{\fehmedian}{\ensuremath{-2.44 \pm 0.09}\xspace}
\newcommand{\fehdisp}{\ensuremath{0.22 ^{+0.10}_{-0.07}}\xspace}

\newcommand{\jcarthreesmall}{\ensuremath{19.9_{-0.9}^{+1.0}}\xspace}
\newcommand{\jcarthreemedium}{\ensuremath{20.1_{-0.9}^{+1.0}}\xspace}
\newcommand{\jcarthreelarge}{\ensuremath{20.2_{-0.9}^{+1.0}}\xspace}

\newcommand{\jcartwosmall}{\ensuremath{17.9_{-0.5}^{+0.6}}\xspace}
\newcommand{\jcartwomedium}{\ensuremath{18.0_{-0.5}^{+0.5}}\xspace}
\newcommand{\jcartwolarge}{\ensuremath{18.2_{-0.5}^{+0.5}}\xspace}
\newcommand{\jcartwoalpha}{\ensuremath{18.1_{-0.5}^{+0.5}}\xspace}

\newcommand{\dcartwosmall}{\ensuremath{16.9_{-0.3}^{+0.3}}\xspace}
\newcommand{\dcartwomedium}{\ensuremath{17.4_{-0.3}^{+0.3}}\xspace}
\newcommand{\dcartwolarge}{\ensuremath{18.0_{-0.4}^{+0.4}}\xspace}
\newcommand{\dcartwoalpha}{\ensuremath{17.1_{-0.3}^{+0.3}}\xspace}

\newcommand{\dcarthreesmall}{\ensuremath{17.8_{-0.5}^{+0.5}}\xspace}
\newcommand{\dcarthreemedium}{\ensuremath{18.3_{-0.5}^{+0.6}}\xspace}
\newcommand{\dcarthreelarge}{\ensuremath{18.8_{-0.7}^{+0.6}}\xspace}

\newcommand{\CarII}{Car~II\xspace}
\newcommand{\CarIII}{Car~III\xspace}
\newcommand{\noinfo}{......\xspace}

\newcommand{\Fermi}{{\it Fermi}\xspace}

\newcommand{\gammaRays}{{$\gamma$ rays}\xspace}
\newcommand{\gammaRayHyph}{{$\gamma$-ray}\xspace}

\newcommand{\irf}[1]{\texttt{#1}\xspace}
\newcommand{\stools}{\emph{ScienceTools}\xspace}
\newcommand{\sigmav}{\ensuremath{\langle \sigma v \rangle}\xspace}

\shorttitle{Spectroscopic Analysis of Carina II \& III}
\shortauthors{Li et~al.}

\begin{document}

\title{Ships passing in the night: spectroscopic analysis of two ultra-faint satellites in the constellation Carina
\altaffilmark{*}\altaffilmark{\dag}\altaffilmark{\ddag}}

\altaffiltext{*}{This paper includes data gathered with the 6.5-meter Magellan Telescopes located at Las Campanas Observatory, Chile.}
\altaffiltext{\dag}{This paper includes data gathered with Anglo-Australian Telescope in Australia.}
\altaffiltext{\ddag}{Based on data products from observations made with ESO Telescopes at the La Silla Paranal Observatory under programme ID 298.B-5027.}


\def\andname{}
\author{
T.~S.~Li\altaffilmark{1,2},
J.~D.~Simon\altaffilmark{3},
A.~B.~Pace\altaffilmark{4,29},
G.~Torrealba\altaffilmark{5},
K.~Kuehn\altaffilmark{6},
A.~Drlica-Wagner\altaffilmark{1},
K.~Bechtol\altaffilmark{7},
A.~K.~Vivas\altaffilmark{8},
R.~P.~van~der~Marel\altaffilmark{9,10},
M.~Wood\altaffilmark{11,12,13},
B.~Yanny\altaffilmark{1},
V.~Belokurov\altaffilmark{14},
P.~Jethwa\altaffilmark{15},
D.~B.~Zucker\altaffilmark{16,17},
G.~Lewis\altaffilmark{18},
R.~Kron\altaffilmark{1,19},
D.~L.~Nidever\altaffilmark{20},
M.~A.~S\'anchez-Conde\altaffilmark{21,22},
A.~P.~Ji\altaffilmark{3,30},
B.~C.~Conn\altaffilmark{23},
D.~J.~James\altaffilmark{24},
N.~F.~Martin\altaffilmark{25,26},
D.~Martinez-Delgado\altaffilmark{27},
N.~E.~D.~No\"el\altaffilmark{28}
\\ \vspace{0.2cm} (MagLiteS Collaboration) \\
}
\affil{$^{1}$ Fermi National Accelerator Laboratory, P.O.\ Box 500, Batavia, IL 60510, USA}
\affil{$^{2}$ Kavli Institute for Cosmological Physics, University of Chicago, Chicago, IL 60637, USA}
\affil{$^{3}$ Observatories of the Carnegie Institution for Science, 813 Santa Barbara St., Pasadena, CA 91101, USA}
\affil{$^{4}$ George P. and Cynthia Woods Mitchell Institute for Fundamental Physics and Astronomy, and Department of Physics and Astronomy, Texas A\&M University, College Station, TX 77843, USA}
\affil{$^{5}$ Institute of Astronomy and Astrophysics, Academia Sinica, P.O. Box 23-141, Taipei 10617,Taiwan}
\affil{$^{6}$ Australian Astronomical Observatory, North Ryde, NSW 2113, Australia}
\affil{$^{7}$ Large Synoptic Survey Telescope, 950 North Cherry Avenue, Tucson, AZ 85721, USA}
\affil{$^{8}$ Cerro Tololo Inter-American Observatory, National Optical Astronomy Observatory, Casilla 603, La Serena, Chile}
\affil{$^{9}$ Space Telescope Science Institute, 3700 San Martin Drive, Baltimore, MD 21218, USA}
\affil{$^{10}$ Center for Astrophysical Sciences, Department of Physics \& Astronomy, Johns Hopkins University, Baltimore, MD 21218, USA}
\affil{$^{11}$ Kavli Institute for Particle Astrophysics \& Cosmology, P.O. Box 2450, Stanford University, Stanford, CA 94305, USA}
\affil{$^{12}$ SLAC National Accelerator Laboratory, Menlo Park, CA 94025, USA}
\affil{$^{13}$ Department of Physics, Stanford University, 382 Via Pueblo Mall, Stanford, CA 94305, USA}
\affil{$^{14}$ Institute of Astronomy, University of Cambridge, Madingley Road, Cambridge CB3 0HA, UK}
\affil{$^{15}$ European Southern Observatory, Karl-Schwarzschild-Str. 2, 85748 Garching, Germany}
\affil{$^{16}$ Department of Physics \& Astronomy, Macquarie University, Sydney, NSW 2109, Australia}
\affil{$^{17}$ Australian Astronomical Observatory, 105 Delhi Rd, North Ryde, NSW 2113, Australia}
\affil{$^{18}$ Sydney Institute for Astronomy, School of Physics, A28, The University of Sydney, NSW 2006, Australia}
\affil{$^{19}$ Department of Astronomy and Astrophysics, University of Chicago, Chicago, IL 60637, USA}
\affil{$^{20}$ National Optical Astronomy Observatory, 950 N. Cherry Ave., Tucson, AZ 85719, USA}
\affil{$^{21}$ Instituto de F\'{\i}sica Te\'orica UAM/CSIC, Universidad Aut\'onoma de Madrid, E-28049 Madrid, Spain}
\affil{$^{22}$ Departamento de F\'isica Te\'orica, M-15, Universidad Aut\'onoma de Madrid, E-28049 Madrid, Spain}
\affil{$^{23}$ Research School of Astronomy \& Astrophysics, Mount Stromlo Observatory, Cotter Road, Weston Creek, ACT 2611, Australia}
\affil{$^{24}$ Event Horizon Telescope, Harvard-Smithsonian Center for Astrophysics, 60 Garden Street, Cambridge, MA 02138, USA}
\affil{$^{25}$ Universit\'e de Strasbourg, CNRS, Observatoire astronomique de Strasbourg, UMR 7550, F-67000 Strasbourg, France}
\affil{$^{26}$ Max-Planck-Institut f\"{u}r Astronomie, K\"{o}nigstuhl 17, D-69117 Heidelberg, Germany}
\affil{$^{27}$ Astronomisches Rechen-Institut, Zentrum f\"ur Astronomie der Universit\"at Heidelberg, M{\"o}nchhofstr. 12-14, 69120 Heidelberg, Germany}
\affil{$^{28}$ Department of Physics, University of Surrey, Guildford, GU2 7XH, UK}
\altaffiliation{$^{29}$ Mitchell Astronomy Fellow}
\altaffiliation{$^{30}$ Hubble Fellow}
\email{tingli@fnal.gov}

\begin{abstract}

We present Magellan/IMACS, Anglo-Australian Telescope/AAOmega+2dF, and  Very Large Telescope/GIRAFFE+FLAMES spectroscopy of the Carina~II (\CarII) \& Carina~III (\CarIII) dwarf galaxy candidates, recently discovered in the Magellanic Satellites Survey (MagLiteS).
We identify 18 member stars in \CarII, including 2 binaries with variable radial velocities and 2 RR Lyrae stars.
The other 14 members have a mean heliocentric velocity $v_{\rm hel} = \vbulk$~\kms and a velocity dispersion of $\sigma_v = \vdisp$~\kms.
Assuming \CarII is in dynamical equilibrium, we derive a total mass within the half-light radius of $\mass$~\msun, indicating a mass-to-light ratio of \masstolight~\msun/\lsun.
From equivalent width measurements of the calcium triplet lines of 9 RGB stars, we derive a mean metallicity of $\feh = \fehmedian$ with dispersion $\sigma_{\feh} = \fehdisp$.
Considering both the kinematic and chemical properties, we conclude that \CarII is a dark-matter-dominated dwarf galaxy.
For \CarIII, we identify 4 member stars, from which we calculate a systemic velocity of $v_{\rm hel} = \vbulkcariii$~\kms. 
The brightest RGB member of \CarIII has a metallicity of $\feh = -1.97 \pm 0.12$. 
Due to the small size of the \CarIII spectroscopic sample, we cannot conclusively determine its nature.
Although these two systems have the smallest known physical separation ($\Delta d\sim10~\kpc$) among Local Group satellites, the large difference in their systemic velocities, $\sim200~\kms$, indicates that they are unlikely to be a bound pair. 
One or both systems are likely associated with the Large Magellanic Cloud (LMC), and may remain LMC satellites today. 
No statistically significant excess of \gammaRayHyph emission is found at the locations of \CarII and \CarIII in eight years of \textit{Fermi}-LAT data.

\end{abstract}


\section{INTRODUCTION}
\label{sec:intro}

\setcounter{footnote}{30}
The standard cosmological model with cold dark matter predicts that structure forms hierarchically over a wide range of size scales. 
The two most prominent satellites of the Milky Way, the Large and Small Magellanic Clouds (LMC and SMC), are both sufficiently massive to expect that they hosted their own populations of luminous satellites prior to their arrival at the Milky Way~\citep{D'Onghia:2008a,Sales:2011a,Dooley2017}. 
Indeed, the spatial distribution of the newly discovered ultra-faint dwarf galaxies in the Dark Energy Survey \citep[DES;][]{Abbott:2005bi} is heavily biased toward the direction of the LMC and SMC, providing strong but indirect observational evidence for the existence of ``satellites of satellites'' around our Milky Way~\citep{koposov15,bechtol15,dw15b,deason15,jethwa16,Sales2017}. 
Motivated by this distinct anisotropy in the southern satellite distribution, the Magellanic Satellites Survey (MagLiteS) is imaging the unexplored area on the other side of the Magellanic Clouds with the Dark Energy Camera~\citep[DECam;][]{flaugher_2015_decam} on the Blanco 4m telescope at Cerro Tololo Inter-American Observatory. 
MagLiteS is described in more detail by \citet{dw16}.

Recently, a pair of dwarf galaxy candidates located on the outskirts of the LMC were discovered using photometric data from MagLiteS: Carina~II (\CarII) and Carina~III (\CarIII) \citep[][hereafter Paper I]{torrealba:18}. 
These two systems are both extremely faint and close to us, with absolute magnitudes of $M_V \sim -4.5$ and $M_{V} \sim -2.4$, and heliocentric distances of $d \sim 37$~kpc and $d\sim 28$~kpc, respectively. 
\CarII ($r_{1/2} \sim 90$~pc) is significantly more extended than \CarIII ($r_{1/2}\sim 30$~pc). 
Remarkably, these two objects form a close pair both on the sky (where they have a projected separation of $18\arcmin$ or $\sim 150$~pc) and along the line of sight (where they are $\sim$10~kpc apart), raising the question of whether \CarII and \CarIII are gravitationally bound. 
Furthermore, due to the proximity of both systems to the LMC ($\sim$20~kpc), it seems likely that one or both are (or were) physically associated with the Magellanic Clouds. 
Kinematic information, such as line-of-sight velocities, is necessary to address these hypotheses, and confirm the nature of the two systems.

Soon after the initial discovery in 2016 December, we began a spectroscopic follow-up program with the Magellan Baade Telescope, the Anglo-Australian Telescope (AAT), and the Very Large Telescope (VLT). 
Rapid follow-up with the AAT and VLT was possible thanks to short turn-around time for approval of service observations and Director's Discretionary time.  
Our multi-pronged observational strategy enabled both deep observations of a smaller number of faint targets with the 8-m-class Magellan and VLT, and wider-field observations of a large number of brighter targets (in both \CarII and \CarIII together) with the AAT.  

Here, we report the first spectroscopic analysis of the \CarII and \CarIII dwarf galaxy candidates discovered in MagLiteS. In \S\ref{sec:observations}, we describe the observations with all three telescopes, and the data reduction procedures.
In \S\ref{sec:results}, we detail the results from our spectroscopic program, including the set of spectroscopic members, and measurements of the radial velocity, velocity dispersion, mean metallicity, and metallicity dispersion for each dwarf galaxy candidate. 
In \S\ref{sec:discussion}, we discuss the implications of these derived parameters as they relate to the classification of \CarII and \CarIII, along with other unique features of this pair -- specifically, the possible tidal interaction between the two systems, and the association of \CarII and \CarIII with the Magellanic Clouds. 
We also briefly discuss the search for dark matter annihilation within the Carina systems. 
We conclude in \S\ref{sec:summary}.

The photometry in this work has been dereddened using the \citet{1998ApJ...500..525S} extinction map around \CarII and \CarIII.  Because of the relatively low Galactic latitude, the average reddening in this region is $E(B-V)\sim0.19$.

\section{OBSERVATIONS AND DATA REDUCTION}
\label{sec:observations}

\begin{deluxetable*}{lcccrccrccrr}
\tablecaption{Observations}
\tablewidth{0pt}
\tablehead{
\colhead{Mask/Run} &
\colhead{$\alpha$ (J2000)} &
\colhead{$\delta$ (J2000)} &
\colhead{Slit PA} &
\colhead{$\lambda$/$\Delta$$\lambda$} &
\colhead{disp.} &
\colhead{\# of} &
\colhead{$\Sigma$$t_{\rm exp}$} &
\colhead{seeing} &
\colhead{MJD\tablenotemark{a}} &
\colhead{\# of} &
\colhead{\# of useful} \\
\colhead{name}&
\colhead{(h$\,$ $\,$m$\,$ $\,$s)} &
\colhead{($^\circ\,$ $\,'\,$ $\,''$)} &
\colhead{(deg)} &
\colhead{} &
\colhead{\AA/pix} &
\colhead{exp} &
\colhead{(sec)} &
\colhead{(\arcsec)} &
\colhead{} &
\colhead{slits/fibers} &
\colhead{spectra}
}
\startdata
\code{IMACS-Car2Mask1}     & 07:36:36.00 & $-$57:57:20.0  & 172.0 & 11,000 & 0.19 & 2 & 4500  & 1\farcs2 & 57779.3 & 72  & 32\\
\code{IMACS-Car2Mask2}     & 07:36:26.00 & $-$58:02:00.0  & 248.0 & 11,000 & 0.19 & 3 & 6000  & 1\farcs5 & 57778.3 & 48  & 33\\
\code{IMACS-Car3Mask1}     & 07:38:28.00 & $-$57:53:30.0  & 317.0 & 11,000 & 0.19 & 6 & 13200 & 2\farcs5 & 57778.2 & 67  & 40\\
\code{IMACS-Car3LongSlit}  & 07:38:34.84 & $-$57:52:11.2  & 180.0 & 11,000 & 0.19 & 1 & 1800  & 1\farcs2 & 57779.3 & 1   &  1\\
\code{AAT-Jan}            & 07:36:33.60 & $-$58:01:12.0  & --    & 10,000 & 0.24 & 5 & 13200 & 1\farcs3 & 57777.6 & 388 & 131\\
\code{AAT-May}             & 07:36:33.60 & $-$58:01:12.0  & --    & 10,000 & 0.24 & 2 & 4800  & 2\farcs0 & 57902.4 & 309 & 55\\
\code{VLT-Feb}             & 07:37:32.40 & $-$57:57:16.0  & --    & 6,000 & 0.20 & 1 & 2775  & 0\farcs3 & 57811.1 & 116 & 116\\[-0.6em]
\enddata
\tablenotetext{a}{The date listed here is the weighted mean observation date over multiple exposures. For \code{AAT-Jan} observations, the date is the weighted mean over multiple nights. 
}
\label{tab:obstable}
\end{deluxetable*}

\begin{figure*}[th!]
\epsscale{1}
\plotone{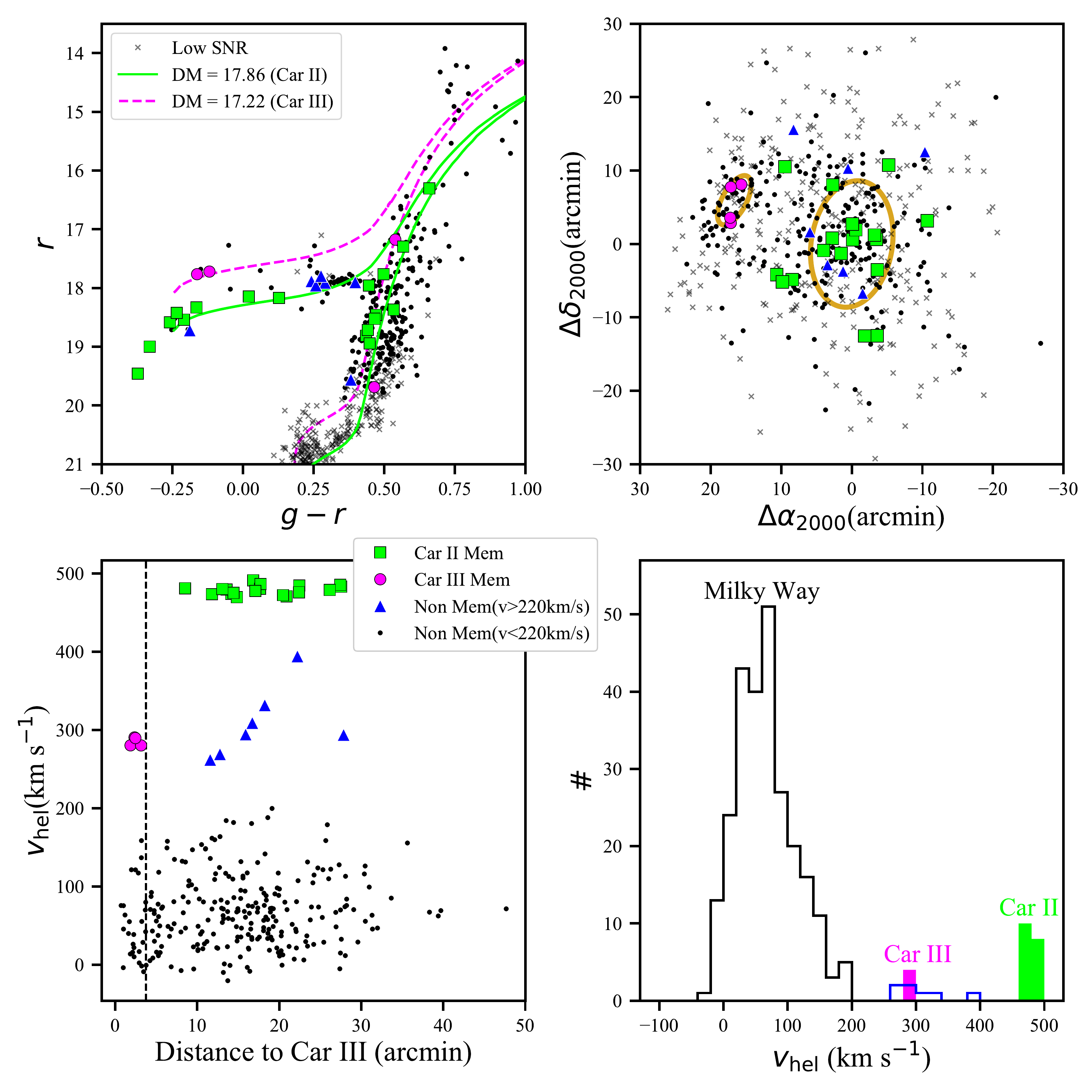}
\caption{\emph{upper-left}: Color-magnitude diagram for observed stars with Magellan/IMACS, AAT/2dF+AAOmega and VLT/GIRAFFE+FLAMES. Overplotted are the PARSEC isochrones of a metal-poor population with age = 12.0 Gyr, [Fe/H] = -2.2 at the distance of \CarII ($m-M = 17.86$, green) and \CarIII ($m-M = 17.22$, magenta).  Green squares indicate 18 members of \CarII, magenta circles indicate 4 members of \CarIII, blue triangles indicate 7 non-members with velocities $v_{hel} > 220\kms$, and small black dots are remaining non-member with velocities $v_{hel} < 220\kms$; black cross markers indicate candidate members that were observed but for which we were unable to obtain velocity measurements (mainly due to low S/N). \emph{upper-right}: Spatial distribution of the targets. Gold ellipses show the half-light radius of \CarII (larger ellipse) and \CarIII (smaller ellipse). \emph{lower-left}: Heliocentric velocity versus distance from the center of \CarIII.  The separation between the \CarII members, \CarIII members and the non-members is obvious. A few non-members have velocities similar to that of \CarIII, but are far away from the \CarIII center. The black dashed line indicates the half-light radius of \CarIII ($r_h=3.75\arcmin$; Paper~I).
\emph{lower-right}: Velocity distribution of 283 stars with successful velocity measurements. \CarII members are indicated as the peak around 480~\kms (in green) and \CarIII members are indicated as the peak around 280~\kms (in magenta). A few stars have velocities close to \CarIII and are shown as a blue histogram. From their position on the sky (upper-right panel) and their distance from \CarIII (lower-left panel) we conclude that they are not members of \CarIII.
}
\label{cmd}
\end{figure*}

\subsection{Magellan/IMACS Spectra}
\label{sec:imacs}

We obtained multi-slit spectroscopy of \CarII and \CarIII with the IMACS spectrograph~\citep{dressler06} on the Magellan Baade telescope on 2017 January 24--25. 
The observing setup was the same as for the spectroscopy of the Tucana~III~\citep{tuc3} and Eridanus~II~\citep{eri2} dwarf galaxies. 
We used the f/4 camera on IMACS, which provides a full field-of-view of $15.4\arcmin \times 15.4\arcmin$.  
The spectrograph was configured with the 1200~$\ell$/mm grating and a tilt angle of 32.4\degr, producing a spectral resolution of $R\sim11,000$ for a 0\farcs7 slit width and a wavelength range of at least $7550-8750$~\AA\ for each slit. 
This wavelength range covers the calcium triplet (CaT) lines around 8500~\AA, used for measuring radial velocities and metallicities of candidate member stars, as well as the telluric absorption lines (Fraunhofer A-band) around 7600~\AA\, used for the correction of velocity errors caused by mis-centering of the stars within the slits (see \citealt{sohn07} for details).

The target selection and mask design for Magellan/IMACS (hereafter IMACS) were performed using the photometry from the original MagLiteS catalog. 
Based on the knowledge of confirmed members of the DES-discovered dwarf galaxies Reticulum~II from~\citet{simon15b} and Tucana~III from~\citet{tuc3}, we used similar selection criteria as described in \citet{tuc3}.
Target selection and mask design used a preliminary estimate for the distance modulus for \CarII (\CarIII) of $m-M = 17.5$ ($m-M = 17.1$). 
For \CarII, the red giant branch (RGB) candidate members were selected to be redder than the fiducial sequence of the metal-poor globular cluster M92 from~\citet{an08}, bluer than a 12~Gyr, $\feh = -2.2$ theoretical PARSEC isochrone\footnote{Note that an earlier version of the PARSEC isochrone was used. See the Appendix for more details regarding the updated PARSEC isochrones.} from~\citet{bressan12}, and brighter than $g=20.8$. 
Several candidate blue horizontal branch (BHB) stars were selected at $17.7 < g < 18.8$ and $g-r < 0.1$; a handful of candidate red horizontal branch (RHB) stars were selected at $17.9 < g < 18.6$ and $0.1 <g-r < 0.3$. 
Potential main sequence turnoff (MSTO) stars were selected using a 0.1 mag wide window in $g - r$ around the PARSEC isochrone for $20.8 < g < 22$ (although because of the observing conditions, we did not get any useful spectra for MSTO candidates).

For \CarIII, we used the same selection criteria as for \CarII, with the exception of shifting all sequences 0.4~mag brighter according to the difference in distance moduli. 
Based on these targets, we designed two slitmasks (\code{IMACS-Car2Mask1} and \code{IMACS-Car2Mask2}) near the center of \CarII and one slitmask (\code{IMACS-Car3Mask1}) near the center of \CarIII, using the \code{maskgen} program\footnote{\url{http://code.obs.carnegiescience.edu/maskgen}}. 
Stars were placed on the slit masks in a category prioritization descending order of BHB, RGB, RHB and MSTO. 
Within each category, priorities were based on brightness and distance from the center of \CarII or \CarIII. Finally, any remaining mask space was filled by stars with photometry that made them unlikely to be members. \code{IMACS-Car2Mask1} contains 72 slits, \code{IMACS-Car2Mask2} contains 48 slits and \code{IMACS-Car3Mask1} contains 67 slits. All the targets, observed with IMACS and other instruments described in later sections, are presented in Figure~\ref{cmd}.

We obtained a 1.7-hr exposure with \code{IMACS-Car2Mask2} and 3.7-hr exposure with \code{IMACS-Car3Mask1} on 2017 January 24, and 1.25-hr exposure on \code{IMACS-Car2Mask1} on 2017 January 25. 
One of the BHB candidates, MAGLITES\,J073834.84$-$575211.2, on \code{IMACS-Car3Mask1} happened to fall in a gap between CCDs, so we also obtained a 30-min exposure on this star with a 0\farcs7-wide long slit (\code{IMACS-Car3LongSlit}) on 2017 January 25.
The observing conditions on both nights were relatively poor, with high humidity and $\sim1\arcsec-3\arcsec$ seeing. 

We reduced the IMACS spectra following the procedures described by~\citet{tuc3} for Tucana~III. 
We first performed the bias subtraction and removal of read-out pattern noise, then we used the Cosmos pipeline~\citep{dressler11,oemler17} to derive an initial wavelength solution and performed the slit mapping, followed by a refined wavelength calibration and spectral extraction using an IMACS pipeline derived from the DEEP2 data reduction pipeline for Keck/DEIMOS \citep{cooper12}. 
For each mask, the extracted spectra from multiple exposures were combined using inverse-variance weighting.
The combined spectra reach a signal-to-noise ratio (S/N) $\sim5$~pixel$^{-1}$ at $r = 19.0$ for \CarII and at $r = 19.3$ for \CarIII.

The details of the instrument setup, observing information, mask information, etc., for IMACS and other instruments described in later sections, are summarized in Table~\ref{tab:obstable}. 

\subsection{AAT/AAOmega+2dF Spectra}
\label{sec:aaomega}
We observed \CarII and \CarIII with the AAOmega Spectrograph~\citep{Sharp2006}, a fiber-fed multi-object spectrograph on the 3.9~m Anglo-Australian Telescope (AAT) at Australian Astronomical Observatory (AAO). 
The AAOmega Spectrograph is fed by the Two Degree Field (``2dF") multi-object system, allowing acquisition of up to 392 simultaneous spectra of objects within a 2\degr\ field on the sky.

AAOmega is a dual-beam spectrograph, which feeds a blue arm and a red arm with a beam splitter at 5700~\AA. 
For the red arm, we utilized the 1700D grating, providing a spectral resolution of $R = 10,000$ and wavelength coverage of $8400-8810$~\AA, which enables us to target the spectral region of the CaT absorption lines for velocity and metallicity measurements. 
For the blue arm, we chose the 580V grating with resolution of $R = 1,300$ and wavelength coverage of $3750-5750$~\AA, which allows us to study additional elements (e.g., carbon) in the blue. 
This paper focuses on the kinematics and metallicities of the Carina systems, and therefore the spectra from the blue arm will be discussed in a future paper.

Observations with AAT/AAOmega+2dF (hereafter AAT) were taken on 2017 January 23 and May 29 through the service observing program, and on 2017 January 25 through classical observing time.  
We obtained three 40-min exposures on January 23,  one 40-min exposure and one 60-min exposure on January 25, and two 40-min exposures on May 29. To ensure accurate velocity determination, the arc frames were taken right before the science exposures at the same position. 
During the January run, the seeing was around 1\arcsec--1\farcs5 with intermittent clouds. 
During the May run, the weather was clear with seeing around 1\farcs6--2\farcs2. 
Among the 392 fibers, 25 of them were assigned to sky positions, 8 of them were assigned to guide stars selected from the UCAC4 catalog~\citep{Zacharias:2013}, and the remaining fibers were assigned to target stars.

The targets for the AAT run were mostly selected using the photometry from the original MagLiteS catalog. 
The RGB and RHB candidates were selected using the best-fit PARSEC isochrone for \CarII ($\mathrm{log~age} = 10.0$, $\feh=-1.7$, $m-M=17.5$) and \CarIII ($\mathrm{log~age} = 9.75$, $\feh=-0.9$, $m-M=17.1$) at the time of the observations\footnote{Note that the final isochrone parameter values reported in Paper~I are different from those used for spectroscopic target selection because the photometry and the fits continued to be refined after the spectroscopic observations were obtained.}. 
The BHB candidates were selected using a fiducial M92 BHB isochrone placed at the distance modulus of \CarII and \CarIII. 

In addition to MagLiteS photometry, we also used photometry from time-series follow-up observations (to search for RR Lyrae stars) acquired with DECam during Blanco 4-m Director's Discretionary and engineering time. 
The exposure times for these follow-up studies were shorter than the original MagLiteS exposures and therefore brighter stars could be observed. 
In addition, $u-$band measurements were performed in the follow-up observations, and a handful of K/M giant candidates were selected based on the $u-g$ and $g-i$ color~\citep[e.g., see Figure 2 of ][]{Yanny2009}. 
Furthermore, we included some RR Lyrae candidates from the preliminary analysis of time-series follow-up studies.
Thanks to the proximity of \CarII and \CarIII on the sky, as well as the large field of view of AAT+2dF, both systems were targeted in a single pointing. 
We assigned RR Lyrae candidates the highest priority, followed by the BHB and K/M giant candidates. 
Stars in two remaining categories (RGB, RHB) were prioritized based on their brightness in $r-$band. 
Note that the target spacing of 2dF is typically 30\arcsec -- 40\arcsec due to fiber collisions, and therefore some targets located close to the centers of \CarII and \CarIII were missed where the target density is high.

The candidate stars were then allocated according to the priorities described above using the fiber configuration program \code{configure}\footnote{\url{https://www.aao.gov.au/science/software/configure}} provided by AAO.  
Flexibility in target allocation with 2dF allowed us to identify bright (S/N $>$ 20 per pixel) non-member stars in the January 23 data, leading to re-allocation of those fibers to alternate targets during the January 25 observations. 
388 candidates were targeted in total from the two nights of observations in the January run; 309 of them were targeted again in the May run.

The data reduction was performed using the \code{2dfdr}\footnote{\url{https://www.aao.gov.au/science/software/2dfdr}} v6.28 data reduction program of the AAO. 
The reduction includes bias subtraction, scattered light subtraction, flat-fielding, optimal spectral extraction, wavelength calibration, sky subtraction, and frame combination with cosmic ray rejection. 
Wavelength calibration was first performed using the arc frames taken immediately before or after each science exposure, followed by a recalibration with a second order polynomial fit using sky emission lines. 
As the observations were taken from different nights, the reduced spectra were corrected for the heliocentric motion of the Sun at each exposure, before the spectra from multiple exposures were combined using inverse-variance weighting. To detect possible binary stars, we combined the spectra from the January run (\code{AAT-Jan}) and May run (\code{AAT-May}) separately for the velocity measurements. 
The combined spectra have S/N $\sim5$~pixel$^{-1}$ at $r = 18.7$ for \code{AAT-Jan} and at $r = 18.0$ for \code{AAT-May}.

\subsection{VLT/GIRAFFE+FLAMES Spectra}
\label{sec:giraffe}

After the observations with AAT and Magellan, we also observed \CarII and \CarIII with the GIRAFFE+FLAMES spectrograph~\citep{pasquini00} on the 8.2-m Kueyen telescope (UT2) based at the ESO-VLT through Director's Discretionary Time. 
Observations were taken in MEDUSA mode, which allows the simultaneous observation of up to 132 objects, with a minimum target separation of 11\arcsec~due to fiber collisions. 
On the night of 2017 February 26, one 2775-second exposure was taken under excellent seeing condition ($\sim 0\farcs3)$. 
The LR8 grating was used for this observation, which covers the wavelength range from 8206 -- 9400~\AA~at a resolution of $R\sim6,000$.  
The calibration frames, including biases, flats and ThAr arcs, were taken at the end of the night.

Target selection for VLT/GIRAFFE+FLAMES (hereafter VLT) was done in a similar way as for the AAT, with the exception that we manually shifted the best-fit PARSEC isochrone $g-r \sim 0.07$ bluer, based on the confirmed \CarII members identified by IMACS and AAT\footnote{The original PARSEC synthetic isochrones used an out-of-date DECam system response. See details about this shift in the Appendix.}. 
As the field of view of FLAMES is about 25\arcmin\ in diameter, we centered the exposure field in between \CarII and \CarIII and we therefore missed some of the BHB members found in the AAT data (see \S\ref{sec:results}). 
116 targets were selected to feed to FLAMES, with 13 fibers assigned to blank sky positions. 

As only a single exposure was obtained with VLT, we first removed cosmic rays using L.A.Cosmic~\citep{vandokkum01}. 
We then reduced the data with the GIRAFFE \code{Gasgano} pipeline (v2.4.8) provided by ESO for bias subtraction, flat-fielding, wavelength calibration and spectral extraction of individual objects. 
We performed a wavelength re-calibration using sky emission lines and a sky subtraction with our own code. 
For details, we refer to the spectroscopic analysis of the Horologium~I dwarf galaxy (Li et al., in prep.). 
In summary, a first-order wavelength correction derived from sky lines was applied to every spectrum to compensate for the wavelength shift likely caused by the temperature changes between the science observing during the night and the calibration frames taken at the end of the night. 
We then combined the 13 sky fibers into a master sky spectrum. 
To compensate for fiber-to-fiber throughput and resolution variations, for each target spectrum we degraded the resolution of either the target spectrum or the master sky spectrum (whichever had the higher resolution) and then scaled the master sky spectrum to match the intensity of the sky lines in the target spectrum before the subtraction.
The final reduced spectra (referred to as \code{VLT-Feb}) have S/N $\sim7$~pixel$^{-1}$ at $r \sim 19.8$.




\section{RESULTS}
\label{sec:results}

In this section, we present the results derived from the observations taken from three telescopes. 
We first determine the radial velocity of each individual candidate star. 
We then identify member stars based on the velocity, spatial location, and location on the color-magnitude diagram. 
After identifying the member stars, we also compute the systemic velocity, velocity dispersion, mean metallicity and metallicity dispersion for \CarII and \CarIII.

We use the distance moduli (dereddened) and structural parameters from Paper~I for the analysis in this work, unless otherwise stated. 
These parameters, together with the derived quantities in this section, are summarized in Table~\ref{tab:car2_table}. 
We note that for \CarII, the distance modulus derived from RR Lyrae in Paper~I has smaller uncertainties and is used in this work.

\begin{deluxetable*}{llrr}
\tablecaption{Summary of Properties of Carina~II and Carina~III$^a$}
\tablewidth{0pt}
\tablehead{
\colhead{Row} & \colhead{Property} & \colhead{Carina~II}  & \colhead{Carina~III}
}
\startdata
(1) & RA (J2000)                            & $114.1066 \pm 0.0070$  & $114.6298 \pm 0.0060$ \\
(2) & Dec (J2000)                           & $-57.9991 \pm 0.0100$  & $-57.8997 \pm 0.0080$ \\
(3) & $(m-M)$                               & $17.86 \pm 0.02^b$     & $17.22 \pm 0.10$      \\
(4) & Heliocentric Distance (kpc)           & $37.4 \pm 0.4^b$       & $27.8 \pm 0.6$        \\
(5) & $M_{V,0}$                             & $-4.5 \pm 0.1$         & $-2.4 \pm 0.2$        \\
(6) & $L_{V,0}$ (\lsun)                     & $5.4 \pm 0.5 \times 10^3$ &  $7.8 ^{+1.6}_{-1.3} \times 10^2$  \\
(7) & $r_{\rm 1/2}$ (arcmin)                & $8.69 \pm 0.75$        & $3.75 \pm 1.00$  \\
(8) & $r_{\rm 1/2}$ (pc)                    & $91 \pm 8$             & $30 \pm 9$            \\
(9) & $\epsilon=1-b/a$                 & $0.34 \pm 0.07$        & $0.55 \pm 0.18$      \\
(10) & PA (N to E; deg)                     & $170 \pm 9$            & $150 \pm 14$          \\ [0.5em]
\hline\\[-0.5em]
(11) & Number of Members 					& 14$^c$				 &		4 		\\
(12)  & $v_{\rm hel}$ (\kms)                & \vbulk                 &    \vbulkcariii      \\
(13)  & $v_{\rm GSR}$ (\kms)                & \vgsr                  &    \vgsrb      \\
(14)  & $\sigma_v$ (\kms)                   & \vdisp                 &     \vdispcariii$^d$      \\
(15)  & $M_{\rm half}$ (\msun)              & \mass                  &     \noinfo       \\
(16)  & $M/L_{V}$ (\msun/\lsun)             & \masstolight           &     \noinfo       \\
(17)  & $\frac{dv}{d\chi}$ (\kms~arcmin$^{-1}$) & $0.0 \pm 0.3$      &     \noinfo       \\
(18)  & Mean metallicity                    & \fehmedian             &      $-1.97^e$      \\
(19)  & Metallicity dispersion (dex)        & \fehdisp               &       \noinfo      \\
(20)  & $\log_{10}{J(0.1\degr)}$ (GeV$^{2}$~cm$^{-5}$) & \jcartwosmall   & \jcarthreesmall$^d$ \\
(21)  & $\log_{10}{J(0.5\degr)}$ (GeV$^{2}$~cm$^{-5}$) & \jcartwolarge   & \jcarthreelarge$^d$ \\
(22)  & $\log_{10}{D(0.1\degr)}$ (GeV~cm$^{-2}$) & \dcartwosmall   & \dcarthreesmall$^d$ \\
(23)  & $\log_{10}{D(0.5\degr)}$ (GeV~cm$^{-2}$) & \dcartwolarge   & \dcarthreelarge$^d$ \\[-0.6em]
\enddata
\tablecomments{
(a) Rows $(1)-(10)$ are taken or derived from Paper I.  Values in rows $(11)-(23)$ are derived using the measurements in this paper. All values reported here (and in this paper) are from the 50th percentile of the posterior probability distributions. The uncertainties are from the 16th and 84th percentiles of the posterior probability distributions.\\
(b) The distance derived from RR Lyrae in Paper I is listed here as the heliocentric distance of Car~II and is used throughout the paper. For other quantities, parameters derived from CMD fit are used instead.\\
(c) There are 18 spectroscopic members but only the 14 non-variable stars are used for kinematic analysis.\\
(d) Note that the velocity dispersion and the J-factor for \CarIII is calculated based on only four spectroscopic members. We caution the use of this calculation as the classification of \CarIII is remain unclear.\\
(e) Note that this is the metallicity of the brightest member in Car III and not the mean metallicity of the system.
}
\label{tab:car2_table}
\end{deluxetable*}

\subsection{Radial Velocity Measurements}
\label{sec:RV}

The reduced spectra from IMACS, AAT, and VLT were used for radial velocity measurements following the method described in~\citet{eri2}. 
We measured the heliocentric radial velocities ($v_\mathrm{hel}$) by fitting the reduced spectra with velocity templates using a Markov chain Monte Carlo (MCMC) sampler and finding the best-fit velocity that maximizes the likelihood defined by Eq. 1 in~\citet{eri2}. 
Instead of using only one velocity template per spectrum, we defined a set of templates for each instrument and used the template that gave the largest likelihood at the best-fit velocity as the best template for each star. 
The template set for each instrument includes at least one metal-rich RGB, one metal-poor RGB, and one BHB star. 
The velocity templates for AAT and IMACS were observed using the same instrument setting as the science observation and were constructed following the description in~\citet{tuc3}. 
We were not able to obtain any velocity template spectra during the VLT run. 
Instead, we used the Keck/DEIMOS templates from \citet{kirby15b}, as the Keck/DEIMOS spectra have a much wider wavelength coverage and a similar resolution ($R\sim6,000)$ as our VLT spectra. 
For the IMACS spectra, we also applied a telluric correction derived using a telluric template to correct for the mis-centering of spectroscopic targets within each slit (see \citealt{eri2} for more details). 

The statistical uncertainty on each velocity measurement is calculated as the standard deviation of the posterior velocity distribution from the MCMC sampler. 
This error is related primarily to the S/N of the spectra, with stellar temperature and metallicity also playing a role. 
Other systematic effects, such as instrument flexure, uncertainties in the wavelength calibration, uncertainties in the template velocity and template mismatching, should also be considered in the final velocity uncertainty budget.  
We estimated the systematic uncertainty as the quadrature difference between repeat measurements and the statistical uncertainty~\citep[c.f.][]{sg07,tuc3,eri2}.
We adopted the systematic floor of 1.0~\kms for IMACS from~\citet{tuc3}. 
For AAT, we determine the systematic floor to be 0.5~\kms using repeat measurements of 18 bright stars (S/N $>8$) from the January run and the May run. 
Since only one exposure was taken with VLT, we were not able to derive a systematic floor with this dataset. 
We adopted a systematic floor of 0.9~\kms from the VLT observations of Horologium~I (Li et al., in prep), which has the same instrument setup as this data set. 
We added these systematic uncertainties in quadrature with the statistical uncertainties to obtain the final reported velocity uncertainties $\delta_v$.

\begin{figure*}[th!]
\epsscale{1}
\plotone{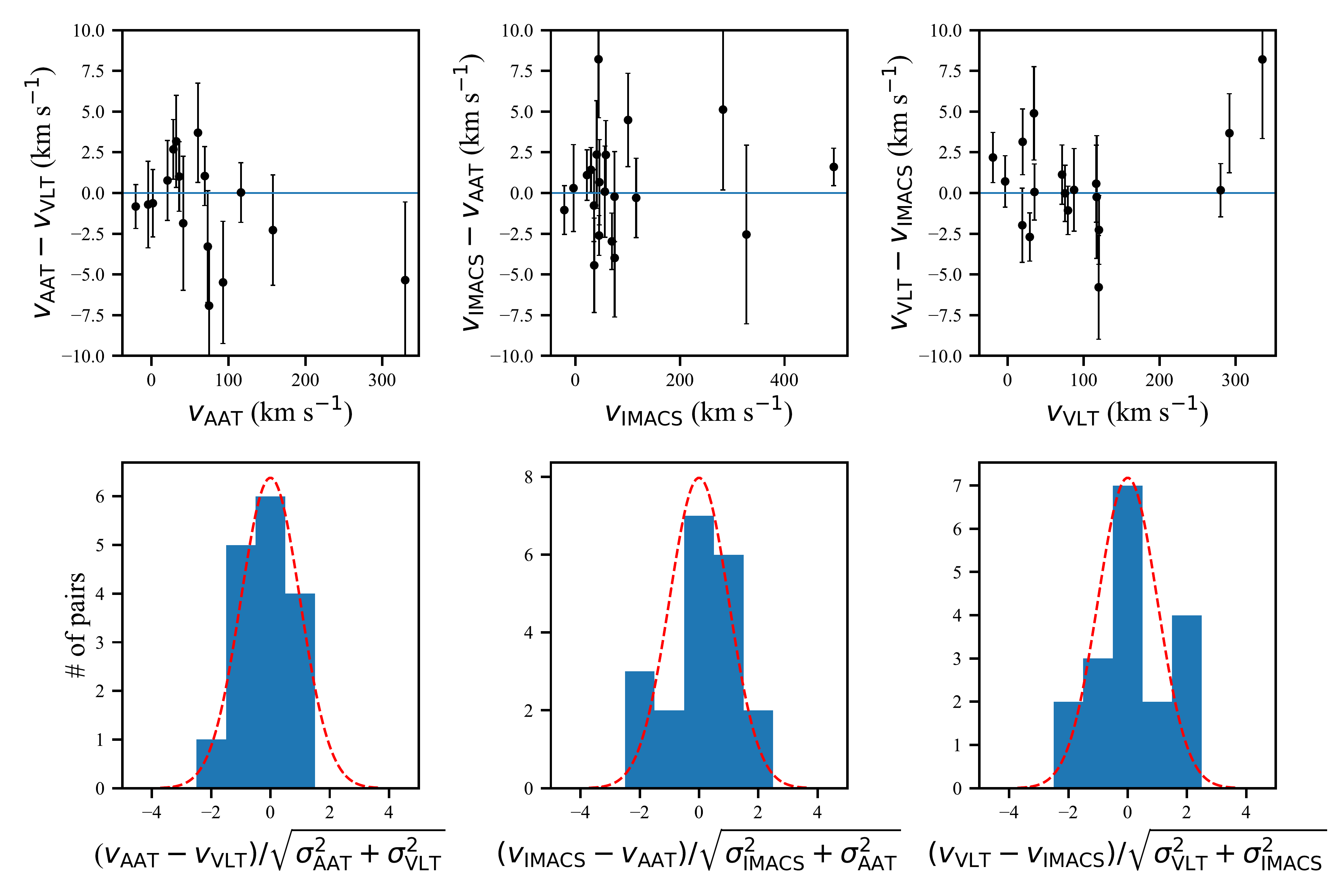}
\caption{$Top~Row$: We compare velocity measurements from different instruments using repeated measurements. There are no obvious zero-point shifts between three instruments. $Bottom~Row$: Radial velocity uncertainty estimation tests using the repeated observations from different instruments. The histograms show the distributions of the velocity difference normalized by the quadrature sum of their uncertainties. The red dashed curves show a normal distribution with zero mean and unit variance scaled by the total number of pairs. The good agreement with the blue histograms indicates that our estimation of the velocity uncertainties is reasonable.
}
\label{rv_compare}
\end{figure*}

In order to combine the velocities derived from three different spectrographs to produce the final dataset for the velocity dispersion determination in \S\ref{sec:vdisp}, we need to verify that there is no systematic offset between these three data sets. 
We compare the repeated measurements from different instruments as shown in the top panels Figure~\ref{rv_compare} and find no obvious systematic offset between any given pair of instruments.  
In order to confirm that our error estimation is reasonable for each instrument, we again use these repeated measurements from each pair of instruments and compute the distribution of velocity differences between the two independent measurements ($v_1$, $v_2$), divided by the quadrature sum of their uncertainties ($\sqrt{\delta_1^2+\delta_2^2}$).
The resulting distributions shown in the bottom panels of Figure~\ref{rv_compare} are well-described by normal distributions with zero mean and unit variance, as shown by red dashed curves in the same plots.  
From this comparison, we conclude that there is no significant zero-point shift between the various spectrographs, and that combining the three datasets will not introduce additional velocity uncertainties. 

In order to study the kinematics of \CarII and \CarIII, as well as the spectroscopic membership in each system, we combined the velocity measurements from the three different instruments into a single data set.  
With this combined sample, we successfully determined the velocities of 283 stars. 
The heliocentric velocities and the associated uncertainties are reported in Table~\ref{tab:car2_spec}. 
We note that although the results reported in the table are from each observing run or each mask, we used the weighted average ($w = 1/\delta^2$) for stars with more than one measurement for the remainder of this paper. 

\subsection{Spectroscopic Membership Determination}
\label{sec:membership}

Figure~\ref{cmd} shows the color-magnitude diagram (CMD), spatial distribution, and velocity distribution of the observed stars in both systems. 
We identified a total of 18 members in \CarII and 4 members in \CarIII from the combined sample (see below).  

The 18 members in \CarII form a coherent velocity peak near 480~\kms in the heliocentric velocity distribution (lower right panel), including 6 BHB members, 2 RR Lyrae members, and 10 RGB members. 
Since the heliocentric velocity of \CarII is quite high relative to the mean velocity and velocity dispersion of the Milky Way halo, there are no foreground contaminants anywhere near the velocity of \CarII. 
Furthermore, all stars within this velocity peak fall on the \CarII isochrone, making the membership of this system unambiguous.
The BHB members are all far from the center of \CarII (r $>$ 10\arcmin) and therefore 4 of them are only observed with AAT. The velocity uncertainties of these BHB members are relatively large as a result of their broad lines and the low S/N of their spectra. 
The brightest RGB member, MAGLITES\,J073621.25$-$575800.3, was observed both in January (AAT \& IMACS) and in May (AAT). The difference between the January and May observations is $\sim$30~\kms. 
Another RGB member, MAGLITES\,J073646.47$-$575910.2, was observed both in January (AAT \& IMACS) and February (VLT), and the difference is $\sim$25~\kms. 
We therefore conclude that these two stars are binaries. 
The 2 RR Lyrae members of \CarII also show velocity variability, and are discussed in more detail in \S\ref{sec:rrl}. 

We find a second narrow peak in the velocity distribution, containing 4 stars, centered at 280~\kms.  All four stars are located within the \CarIII half-light radius and we therefore identify them as \CarIII members. 
While it is difficult to confirm an association based on only four stars, two of the four are BHB stars at the distance of \CarIII ($m-M$ = 17.22, see upper-left panel in Figure~\ref{cmd}).  
The brighter of the two RGB stars lies exactly on the expected \CarIII isochrone, while the fainter one is slightly redder than expected.  
Nevertheless, the combination of the spatial coincidence between these stars and their position in the CMD strongly suggests that this group is related to \CarIII.

Finally, seven candidate stars have velocities in the range of 260--400~\kms and are displayed in blue in Figure~\ref{cmd}.  Given their velocities, these stars are clearly not members of \CarII.  
Their CMD positions and large distance away from \CarIII also indicate that they are not \CarIII members. We used the Besan\c{c}on Galactic stellar model \citep{besancon} to estimate the expected number of foreground Milky Way stars in our spectroscopic sample. 
We selected simulated stars within 0.2 mag of the PARSEC isochrone and with $r < 19.5$. 
We found that in an area of 1~deg$^2$ centered on \CarII there are $\sim$15 simulated stars that have a velocity larger than 260\kms, with a majority at $g-r < 0.4$ (i.e., foreground main-sequence stars). 
The surface density of non-\CarII and \CarIII members in our spectroscopic sample is similar to this value. 
The small number of contaminants from the Besan\c{c}on model further supports the conclusion that the two peaks are associated with \CarII and \CarIII members. 
Given that \CarII is relatively close to the LMC on the sky, some of these non-member stars might also belong to the LMC, which is discussed further in \S\ref{sec:lmcstar}. 

\subsubsection{LMC Contamination}
\label{sec:lmcstar}
The field of \CarII and \CarIII is located 18\degr~($\sim$20 kpc) from the center of the LMC. 
While the visible body of the LMC is contained within the central $\sim 10\degr$~\citep[e.g.,][]{Besla2016}, stars associated with the LMC have been detected as far out as $\sim 20\degr$~\citep[e.g.,][]{Nidever2017}. 
Recently, \citet{Belokurov2016} also reported the detection of a small number of BHB candidates likely associated with the Magellanic Clouds, at a wide range of angular distances, extending out to $\sim$30\degr\ or perhaps even $\sim$50\degr\ from the LMC. 
Interestingly, the diffuse cloud of BHB-like stars appears rather clumpy; the authors identify at least four individual
stream-like structures. 
The most significant of these, the so-called S1 stream, can be traced securely to $20-25$\degr\ from the LMC. 
Further support for the picture in which the LMC is enshrouded in a thin veil of stellar debris comes from the studies of \citet{Mackey2016} and
\citet{Belokurov2017}, who use main sequence and RR Lyrae stars, respectively, to trace a halo-like component around the LMC out to $\sim20$\degr\ from its center. 
Finally, as \citet{Boubert2017} demonstrate using a combination of a stellar evolution code and N-body simulations of the LMC in-fall, the Cloud ought to be surrounded by an envelope of runaway stars. 
These high-velocity escapees are kicked out of the dwarf's disk during stellar binary disruption as a result of core-collapse supernova explosions, and can travel many tens of kpc away from the LMC in all directions. 

It is therefore possible that some stars in our spectroscopic sample belong to the LMC. 
To calculate what the velocities of such stars might be, we used the rotating disk models of~\citet{vanderMarel2016}. 
These imply that the line-of-sight velocity of the LMC disk at the sky position of \CarII and \CarIII is 380~\kms. 
This is $118$~\kms higher than the systemic velocity of the LMC center of mass, due to the fact that far from the center of the galaxy a significant component of its large transverse velocity vector projects along the line of sight. 
The location of \CarII and \CarIII is on the near side of the inclined LMC disk, so the distance to the disk there is only $43.5$ kpc. 
Since the positions of \CarII and \CarIII are near the kinematic minor axis, a possible non-rotating LMC halo population would have more-or-less the same velocity (namely, 380~\kms) as the rotating disk. 
Old populations in the visible part of the LMC have velocity dispersions in the range 20--30~\kms~\citep{vanderMarel2009}. \citet{vdm02} also shows that the velocity dispersion is almost a constant of $\sim20~\kms$ between 2~\kpc and 9~\kpc from the LMC center.
We expect the dispersion at the position of \CarII and \CarIII ($\sim$20 kpc from LMC) to be similar, though it largely depends on the mass and extent of the LMC's dark halo. 
Moreover, the tidal radius of the LMC is $24.0\degr\pm 5.6\degr$~\citep{vanderMarel2014}. 
Therefore, tidal perturbations could affect both the mean velocity and velocity dispersion of LMC stars at the position of \CarII and \CarIII.

The mean velocities inferred here for \CarII and \CarIII are offset by $\sim \pm 100$~\kms, respectively, from the predicted velocities of LMC members. 
Therefore, contamination by LMC members at the \CarII and \CarIII velocities is expected to be negligible. 
We do detect 7 non-member stars (triangles in Figure~\ref{cmd}; see also Table~\ref{tab:car2_spec}) with $v_{\rm hel}$ in the range 260--400~\kms. 
These velocities correspond to much smaller velocities in the Galactocentric frame ($\sim 40$--180~\kms), since \CarII and \CarIII are located almost opposite from the direction of solar motion. 
Among these 7 stars, 6 have $ 0.2 < g-r < 0.4$ and are very likely to be foreground halo stars at much closer distances. 
The seventh star, MAGLITES\,J073634.86$-$580340.6, is a BHB star with $g-r\sim -0.2$. 
From the CMD (see upper-left panel in Figure~\ref{cmd}), its distance is slightly farther than that of \CarII. 
Comparing its $r-$band magnitude with the BHB members in \CarII and \CarIII, we estimate the distance modulus of this star to be $m-M = 18.1\pm0.1$, corresponding to a heliocentric distance of $42\pm 2~$kpc, matching well with the model prediction of $\sim 43.5~$kpc for the near side of the LMC mentioned above. 
This BHB star has independent observations from AAT, IMACS and VLT. 
The weighted average velocity is $v_\mathrm{hel} =331.7\pm2.0~\kms$, showing no evidence of binary motion. 
For comparison, \citet{Munoz2006} detect a group of LMC stars in the field of the Carina dwarf spheroidal galaxy ($\sim22$\degr\ from LMC center and $\sim10$\degr\ from \CarII and \CarIII) with an average radial velocity around $v_\mathrm{hel} = 332~\kms$. 
Therefore, the distance and velocity of this BHB star both suggest an association with the LMC. 
It lies about 18\degr\ from the center of the LMC, making it one of the LMC's most distant spectroscopically confirmed BHB members. 
Clearly, finding additional LMC stars with similar radial velocity at separate positions on the sky would help our understanding of the structure and dynamics of the LMC's outer regions.

\subsubsection{RR Lyrae Stars}
\label{sec:rrl}

\begin{figure}[htb!]
\plotone{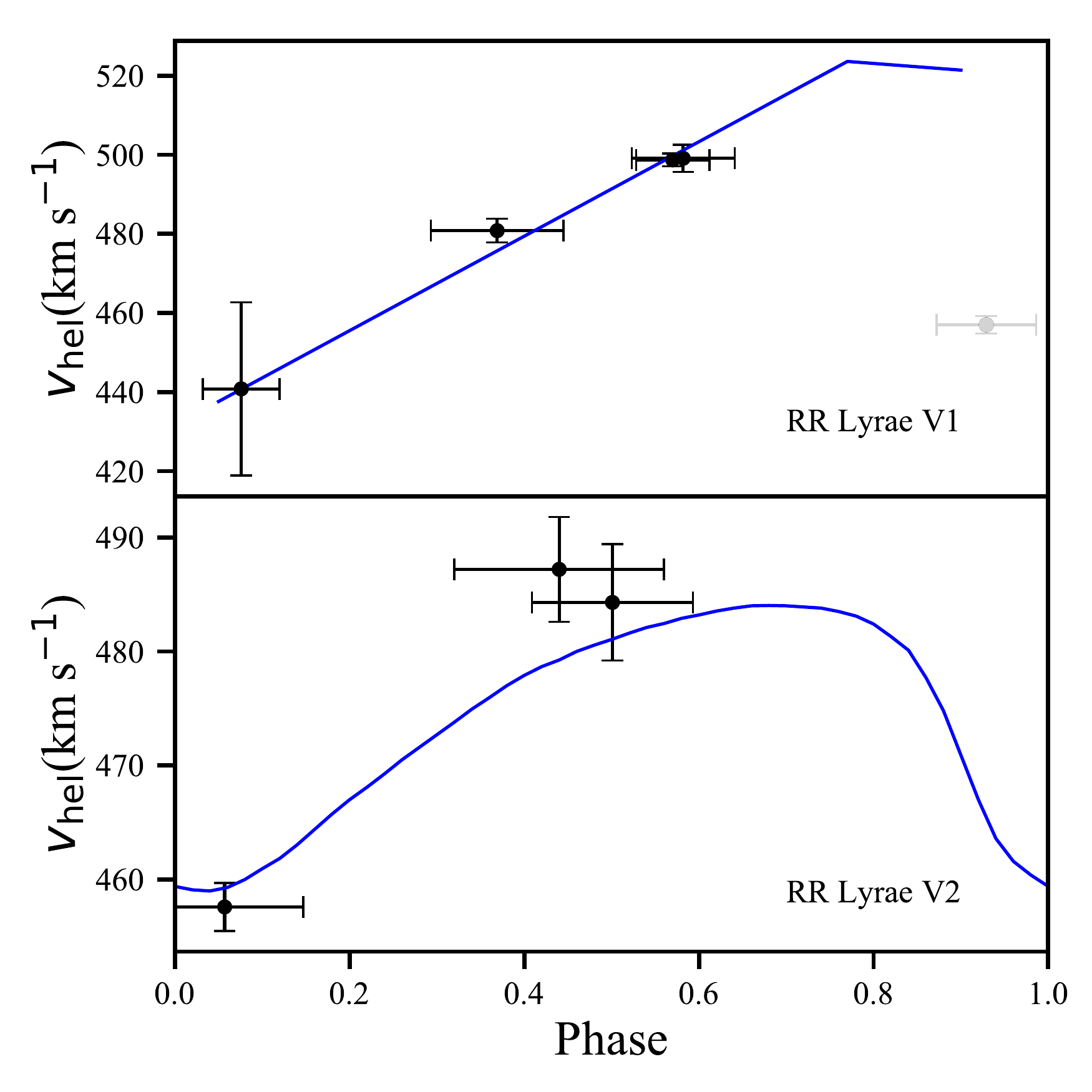}
\caption{Radial velocity curve fits for the observations of the RR Lyrae stars MAGLITES\,J073637.00$-$580114.5 (V1, top) and MAGLITES\,J073645.86$-$575154.1 (V2, bottom). 
Circles represent the measurements from each individual spectrum, which were obtained at different phases during the pulsation cycle. 
The rightmost symbol (grey) for V1 (at phase 0.93) is shown only for reference but it was not used in the fitting of the radial velocity curve since it is located near a large discontinuity in the radial velocity curve. Error bars on the horizontal axis are not real errors but represent the time span (in units of phase) over which each spectrum was obtained. 
Solid lines represent the models fitted to both stars, X~Ari in the case of the RRab star V1, and a template based on T~Sex and DH~Peg in the case of the RRc star V2. }
\label{fig-velRR}
\end{figure}

\CarII contains 3 RR Lyrae stars (Paper~I). 
Our spectroscopic runs targeted 2 of those stars, namely MAGLITES\,J073637.00$-$580114.5 and MAGLITES\,J073645.86$-$575154.1 (or V1 and V2 in Paper~I), which are the two innermost of the RR Lyrae stars, with distances from the center of \CarII of 2\farcm1 and 7\farcm7.
The derivation of the center-of-mass velocity, or systemic velocity, of RR Lyraes requires special treatment since the radial velocity for these stars changes significantly (up to $\sim 100$ km/s for RRab stars) during their pulsation cycle. 
A model of a radial velocity curve must be fitted to the spectroscopic data.
To do this, we followed the procedure developed in \citet{vivas08}. 
The observational data and the fitted model for each star are shown in Figure~\ref{fig-velRR}. 
As the period of RR Lyrae is usually less than a day, we measured the velocities from the January 23 and January 25 AAT observations separately. 
For other measurements, we used the velocity measured over 1-3 combined exposures from that night, as reported in Table~\ref{tab:car2_spec}. 
We therefore have 5 independent velocity measurements for V1 and 3 for V2.

For V1, an RRab star, we used the radial velocity model of the star X Arietis, which was parametrized by \citet{layden94} based on observations made by \citet{oke66}. 
The radial velocity curve of the ab-type RR Lyrae stars (the type of our star V1) has a large discontinuity near maximum light. 
Thus, it is advisable not to measure radial velocities near that phase in the pulsation cycle ($\phi<0.05$ or $\phi>0.9$).
However, at the time of the spectroscopic observations we had not obtained final light curves of the RR Lyrae stars and thus, spectra were taken at random phases. 
Phases were calculated later once the light curves were characterized by Paper~I. 
One of our observations for V1 was indeed not useful since the spectrum was acquired at phase $\phi=0.93$. 
Thus, the radial velocity curve was fitted using 4 observations with phases ranging from 0.08 to 0.58. As seen in Figure~\ref{fig-velRR}, all the individual observations follow very nicely the  model of X Arietis, which was shifted in velocity to match the observations. 
The best match was produced when the systemic velocity was 491~\kms. 
The rms of the fit is 2.8~\kms. 
However, to obtain a more realistic error we followed \citet{vivas05} and included uncertainties due to star-to-star variations of the amplitude of the radial velocity curve as well as possible differences on the exact phase of the systemic velocity. 
We determine a final systemic (heliocentric) velocity for V1 of $491\pm7$~\kms.

For V2, which is an RRc star, we used a template constructed by \citet{duffau06} based on observations of T~Sex and DH~Peg. 
The amplitude of the radial velocity curve of RRc stars is not as large as for RRab variables, nor is there a discontinuity at maximum light. 
Thus, all 3 spectra available for this star can be used to determine its velocity. 
We measure a heliocentric velocity for V2 of $474 \pm 5$~\kms.

The systemic radial velocities obtained for these two RR Lyrae stars confirm that they are members of \CarII.

\subsection{Velocity Dispersion}
\label{sec:vdisp}

\begin{figure*}[th]
\plottwo{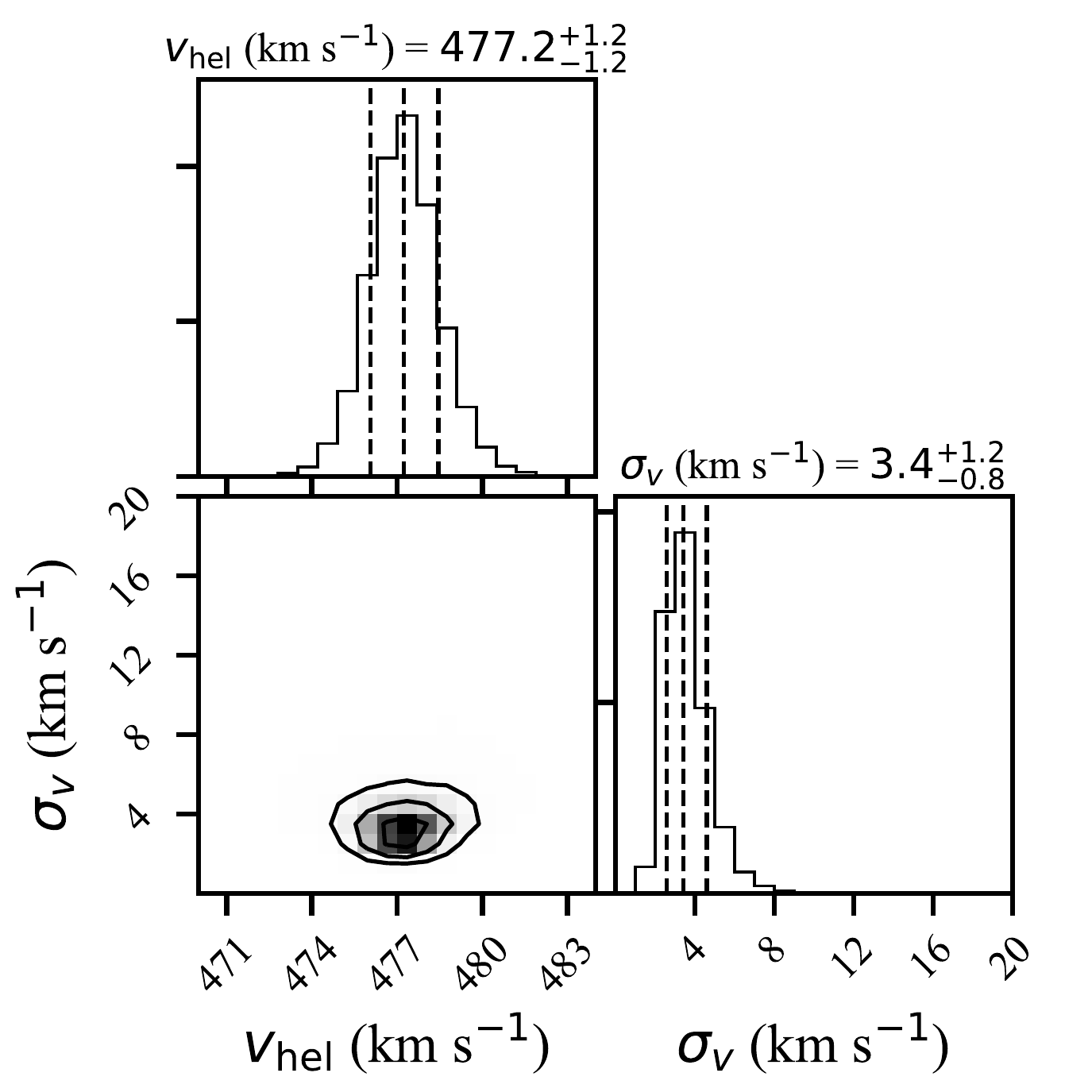}{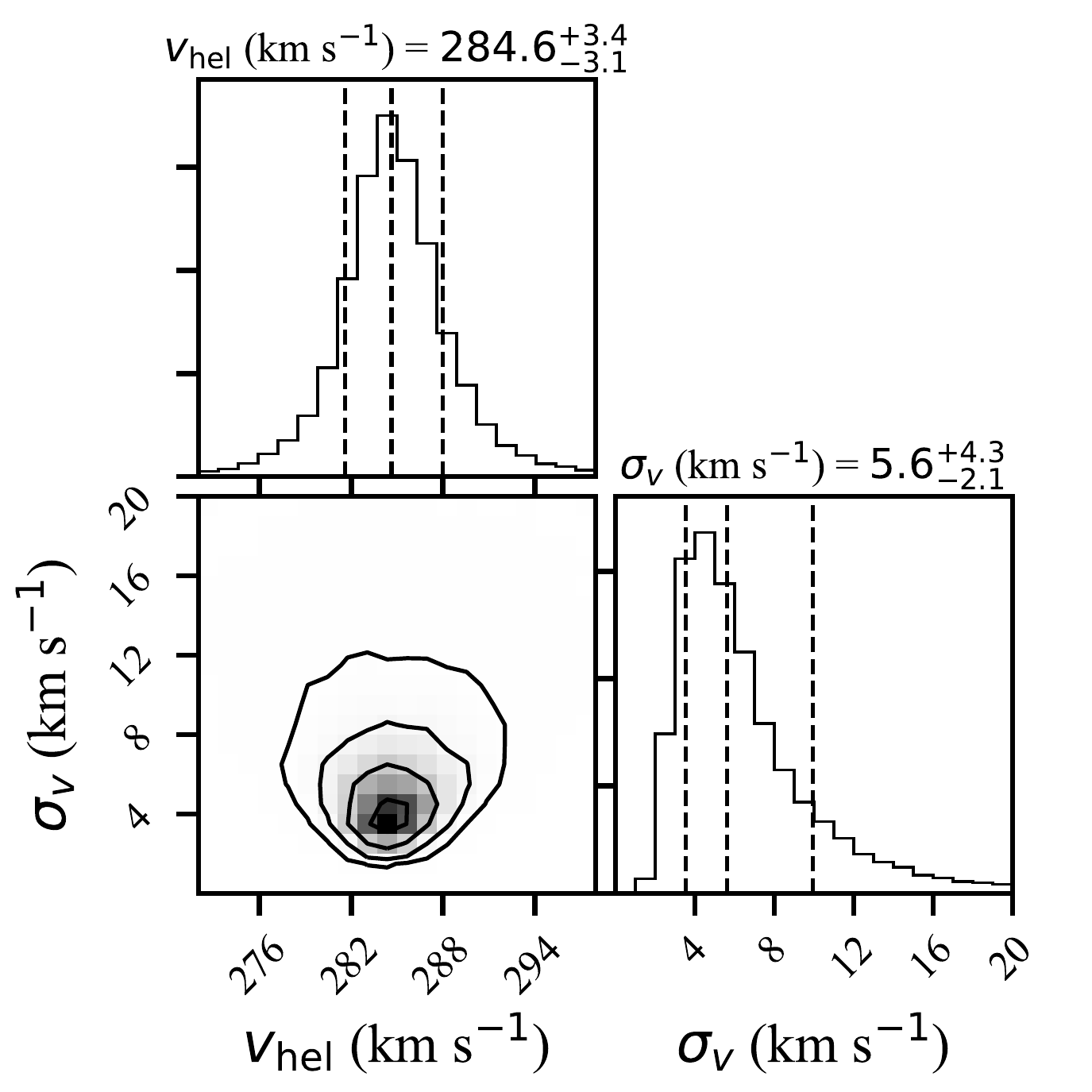}
\caption{Two-dimensional and marginalized posterior probability distributions from an MCMC sampler using a likelihood model for the systemic velocity and velocity dispersion of \CarII \emph{(left)} and those of \CarIII \emph{(right)}.  The 16th, 50th, and 84th percentiles are indicated by dashed lines in the 1-D histograms.}
\label{mcmc}
\end{figure*}

We used 8 RGB stars (excluding the 2 binaries mentioned in \S~\ref{sec:membership}) and 6 BHB stars (hereafter the 14 star sample) to calculate the systemic velocity and the velocity dispersion of \CarII using the 2-parameter Gaussian likelihood function defined in~\citet{Walker06} and an MCMC to sample the distributions of the systemic velocity $v_\mathrm{hel}$ and the velocity dispersion $\sigma_v$. 
We used a flat prior for the systemic velocity with range (455,495)~\kms and a non-informative Jeffreys prior for the velocity dispersion with range (0.01, 100)~\kms (or equivalent to a flat prior in log($\sigma_v$) space with range (-2, 2)). 
The probability distribution from the MCMC is shown in Figure~\ref{mcmc}. 
We find a systemic velocity of $v_{\rm hel} = \vbulk$~\kms and a velocity dispersion of $\sigma_{v} = \vdisp$~\kms, where we report the median of the posterior and the uncertainty calculated from the 16th and 84th percentiles. 

\begin{deluxetable*}{lcccc}
\setlength{\tabcolsep}{0.2in}
\tabletypesize{\scriptsize}
\tablecolumns{5}
\tablewidth{0pc}
\tablecaption{
Systemic velocity and velocity dispersion with different datasets
\label{tab:vdisp_table}
}
\tablehead{
\colhead{Name} & \colhead{Dataset} & \colhead{prior} & \colhead{$v_\mathrm{hel}$} & \colhead{$\sigma_v$}\\
 &  &  & $(\kms)$ & $(\kms)$
}
\startdata  	
14 star sample (default)  &  8 RGB + 6 BHB   & Jeffreys     &  \vbulk          &  \vdisp \\
14 star sample  &  8 RGB + 6 BHB             & flat         &  $477.2\pm1.3$   &  $3.8^{+1.3}_{-0.9}$     \\
16 star sample  &  8 RGB + 6 BHB + 2 RRL     & Jeffreys     &  $477.4\pm1.2$   &  $3.5^{+1.2}_{-0.9}$       \\
RGB only        &  8 RGB                     & Jeffreys     &  $476.6\pm1.2$   &  $3.5^{+1.3}_{-0.9}$       \\
BHB only        &  6 BHB                     & Jeffreys     &  $478.8\pm2.2$   &  $0.9^{+3.7}_{-0.9}$       \\
IMACS only      &  4 RGB                     & Jeffreys     &  $475.8\pm2.1$   &  $4.0^{+2.7}_{-1.7}$       \\
VLT only        &  4 RGB + 2 BHB             & Jeffreys     &  $477.1\pm1.5$   &  $3.3^{+1.7}_{-1.1}$       \\
AAT only        &  5 BHB                     & Jeffreys     &  $480.9\pm2.5$   &  $0.2^{+2.4}_{-0.2}$       \\
one-epoch   & 10 RGB (incl. 2 Binary) +  6 BHB  & Jeffreys   &  $477.8\pm2.2$  & $7.7^{+1.8}_{-1.4}$\\[-0.6em]
\enddata
\tablecomments{
All values reported here (and in this paper) are from the 50th percentile of the posterior probability distributions. The uncertainties are from the 16th and 84th percentiles of the posterior probability distributions.
}
\end{deluxetable*}

In order to test the effects of our input assumptions, we also calculated the systemic velocity and velocity dispersion with different priors and different datasets. 
A summary of these comparisons is presented in Table~\ref{tab:vdisp_table}. 
With a flat prior for velocity dispersion and the same 14 star sample, the velocity dispersion is $3.8_{-0.9}^{+1.3}~\kms$, which is slightly higher than the value determined using the Jeffreys prior with same dataset. 
This result is similar to what has been seen in~\citet{kim15_peg3}. 
If we expand our sample to 16 stars by including the two RR Lyrae stars using the velocities derived in \S\ref{sec:rrl}, both the systemic velocity and the velocity dispersion are very similar to what we determined with the 14 star sample (default sample). 
We then run similar calculations with only RGB members and only BHB members. 
The result with 8 RGB members is again very similar to our default sample, suggesting that the results are mostly constrained by the RGB members (which have smaller velocity uncertainties).  
The 6 BHB members give a smaller dispersion, but with larger uncertainties so that the results are statistically consistent, mainly due to the large velocity uncertainties ($\delta_v\gtrsim4$~\kms) on the BHB stars.

We also calculated the systemic velocity and velocity dispersion using the results from each instrument to see if there is any instrumental bias. 
We obtain very consistent results using the VLT data or IMACS data alone, while the AAT data show much smaller dispersion as the members found by AAT are mostly BHBs (plus RR Lyraes and binaries). 
We additionally calculated the velocity dispersion using the velocity measurements from only one epoch. 
We include the 14 stars in addition to the two binaries. 
For each star the measurement with highest S/N was chosen. 
The derived velocity dispersion is more than doubled compared to that derived from the 14 star sample. 
This exercise mimics a case in which only single-epoch velocity measurements are made for each star and therefore no binary information is available. 
Because of the large velocity amplitudes of the two binary stars, observations made near the velocity extrema of the binary orbits can substantially inflate the apparent velocity dispersion of \CarII. 

Finally, we performed a jackknife test~\citep{macqueen1967} to assess the robustness of the measured velocity dispersion with the 14 star sample, in particular, to check whether the results are driven by any single star.
We remove one star out of the 14 star sample and recompute the average velocity and velocity dispersion. 
In the jackknife runs, the average velocity had a median difference of 0.0~\kms, a standard deviation of 0.3~\kms and a minimum and maximum difference of $-$0.5~\kms and 0.6~\kms. 
For the velocity dispersion the median difference was 0.1~\kms, the standard deviation 0.2~\kms, and the minimum and maximum $-$0.5~\kms and 0.3~\kms. 
We conclude that, apart from the binaries, there are no individual stars whose inclusion or exclusion from the sample significantly affects the kinematics of \CarII. 

We checked if \CarII contains a velocity gradient following the method in \citet{eri2}. 
We calculate a best-fit velocity gradient of $0.0\pm0.3\kms~{\rm arcmin}^{-1}$, consistent with the null model. 
We computed the Bayes Factor comparing the velocity gradient and constant velocity dispersion models and find $\ln{\rm B}=-2.2$, which favors the constant velocity dispersion model (we follow \citealt{2017MNRAS.465.2420W} to interpret the Bayes' Factor value).  
We conclude that there is no evidence for a velocity gradient in \CarII.

For \CarIII, we determined a systemic velocity of \vbulkcariii~\kms and velocity dispersion of \vdispcariii~\kms using the 4 identified members. 
We caution that the small number of stars may not lead to a reliable estimate for the velocity dispersion of \CarIII. 
Furthermore, a single binary star can easily inflate the velocity dispersion. 
We give a more detailed discussion of the implications of the measurements and the nature of \CarIII in \S\ref{sec:properties}.

\subsection{Metallicity and Metallicity Dispersion}
\label{sec:feh}

We measured the metallicity of the red giant member stars in both systems using the equivalent widths (EWs) of the CaT lines. 
Following the procedure described by~\citet{simon15b} and \citet{eri2}, we fitted all three of the CaT lines with a Gaussian plus Lorentzian function and then converted the summed EWs of the three CaT lines to metallicity using the calibration relation from~\citet{carrera13} with absolute $V$ magnitude. 
We first performed the color-transformation from DES-$g$ and DES-$r$ to apparent $V$ magnitude using Equation (5) in~\citet{bechtol15} and then adopted distance moduli of $(m-M) = 17.86$ for \CarII members and $(m-M) = 17.22$ for \CarIII members to calculate absolute magnitudes. 
The statistical uncertainties on the EWs are calculated from the Gaussian and Lorentzian fit. 
We added a systematic uncertainty of 0.2~\AA\ (as determined in \citealt{eri2}) in quadrature with the statistical uncertainties to obtain the final EW uncertainties. 
The metallicity uncertainties shown in Table~\ref{tab:car2_spec} are dominated by the uncertainties on the CaT EWs, with small contributions from the uncertainties on the distances, the stellar photometry, and the uncertainties on the calibration parameters from~\citet{carrera13}.

\begin{figure}[th]
\plotone{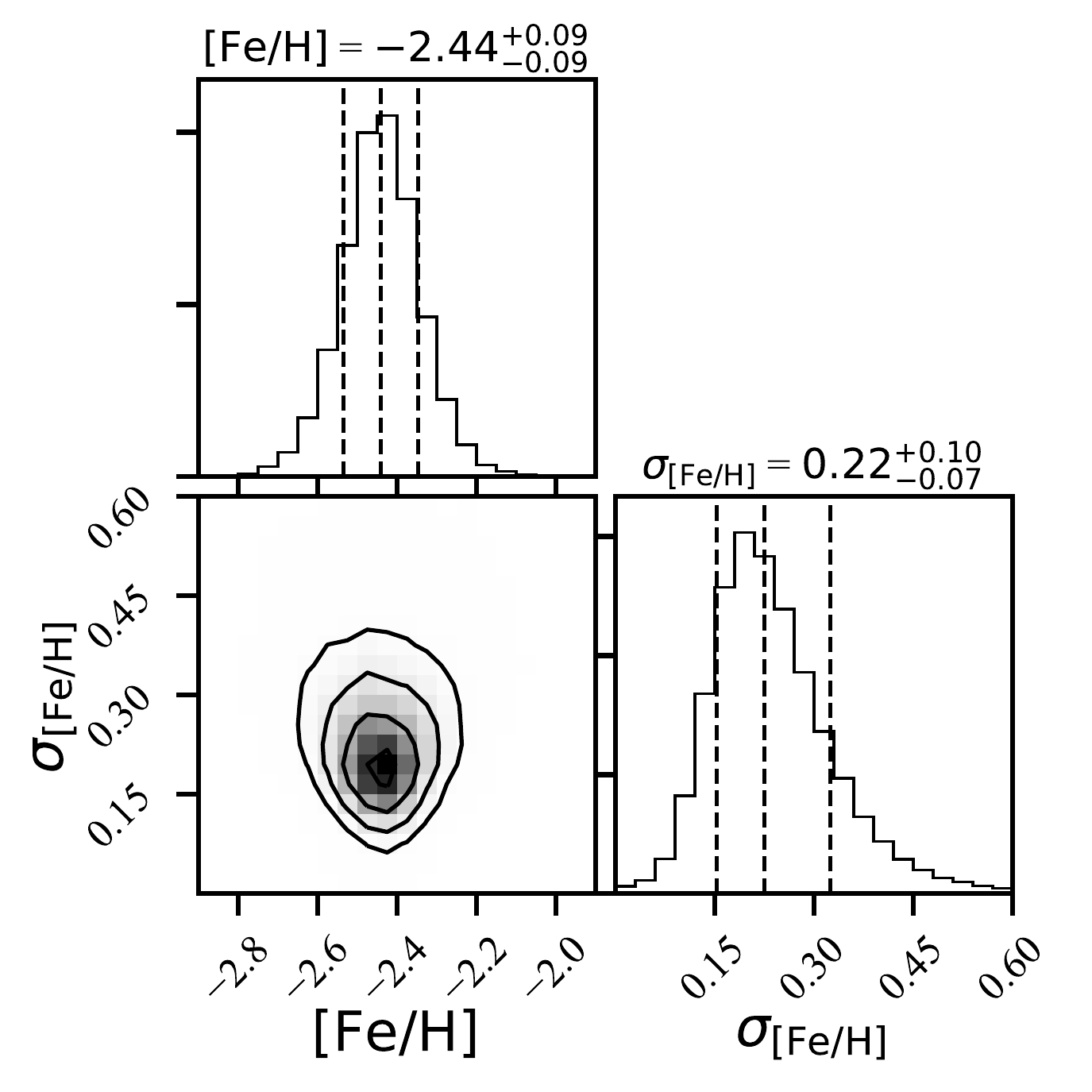}
\caption{Two-dimensional and marginalized posterior probability distributions from an MCMC sampler using a likelihood model for the mean metallicity and metallicity dispersion of \CarII.  The 16th, 50th, and 84th percentiles are indicated by dashed lines in the 1-D histograms.}
\label{mcmc_feh}
\end{figure}

Among the 10 confirmed spectroscopic RGB members in \CarII, we successfully measured metallicities for 9 stars. 
The metallicities of the \CarII members range from $\feh = -2.7$ to $\feh = -1.9$. 
We used a Gaussian likelihood model as described above for the velocities to calculate the mean metallicity and metallicity dispersion of \CarII.  
We find a mean metallicity of $\feh = \fehmedian$, with a dispersion of $\sigma_{\feh} = \fehdisp$. 
The probability distributuion from the MCMC is shown in Figure~\ref{mcmc_feh}. 

For \CarIII, we measured the metallicity of the brightest RGB member, MAGLITES\,J073834.94$-$575705.4, and obtained $\feh = -1.97\pm0.12$ for this star. 
If this RGB member represents the mean metallicity of \CarIII, then the metallicity of \CarII and \CarIII are different at 3-$\sigma$ level.
Note that in Paper~I we obtained $\feh = -1.8\pm0.1$ for \CarII and $\feh = -1.8\pm0.2$ for \CarIII from the isochrone fitting using photometry alone. 
While this metallicity estimate for \CarIII is consistent with that of the brightest RGB member from the spectroscopic measurements, for \CarII the metallicity derived from isochrone fitting is more metal-rich than its spectroscopic mean metallicity.

\section{Discussion}
\label{sec:discussion}

\subsection{Properties of \CarII and \CarIII and Their Possible Association}
\label{sec:properties}

We calculated the mass of \CarII contained within the half-light radius according to the mass estimator from \citet{wolf10} \citep[see also][]{Walker2009}, using the velocity dispersion determined in \S\ref{sec:vdisp} and the half-light radius of \CarII from Paper~I. We found a dynamical mass of  $M_{\rm 1/2} = \mass$~\msun and a mass-to-light ratio of \masstolight~\msun/\lsun for \CarII. 
The reported uncertainties on the dynamical mass and mass-to-light ratio include the uncertainties on the velocity dispersion from this paper, and the uncertainties on the half-light radius and luminosity from Paper~I. 
The mass of \CarII is much larger than its stellar mass, and the mass-to-light ratio is similar to those of other dwarf galaxies with comparable luminosities.
The low average metallicity (\fehmedian) and large metallicity dispersion (\fehdisp) also match observations of other dwarf galaxies with similar luminosities~\citep{kirby13b}.
We therefore conclude that \CarII is a dark matter-dominated dwarf galaxy. 

Because we have only identified 4 members of \CarIII, neither its mass nor its metallicity distribution is significantly constrained. 
We therefore cannot determine whether \CarIII is a dwarf galaxy.
If the metallicity of the brightest confirmed member star ($\feh = -1.97$) represents the average metallicity of the system, then \CarIII is more metal-rich than most of the dwarf galaxies with similar luminosities, but still much more metal-poor than all known star clusters at a similar luminosity.
If the velocity dispersion calculated from the 4 confirmed members is close to the true dispersion of the system, then \CarIII is likely to be a dark matter-dominated dwarf galaxy. 
Given the small sample, though, a single binary star could easily inflate the velocity dispersion, and therefore this dispersion shall be used with caution. 
Interestingly, among the 4 member stars, MAGLITES\,J073834.94$-$575705.4, the brightest RGB member, was observed in both January (IMACS) and February (VLT), and MAGLITES\,J073835.54$-$575622.3, a BHB member, was observed in January (IMACS), February (VLT) and May (AAT). The differences in velocities are consistent with the measurement uncertainties (see Table~\ref{tab:car2_spec}) and therefore we do not see any strong evidence for binarity of these two stars from our observations across 1-3 month baselines. 
However, the velocities of these two stars are about 8~\kms apart (and are the source of the large velocity dispersion on \CarIII). 
This large difference, if indeed not caused by binary motion, could provide a hint that \CarIII should be classified as a dwarf galaxy. 
Identifying more members with deeper observations will be necessary to confirm the nature of \CarIII. 
Observing these bright members at one or two additional epochs will also help determine whether or not they are in binary systems. 

\begin{figure}[th!]
\plotone{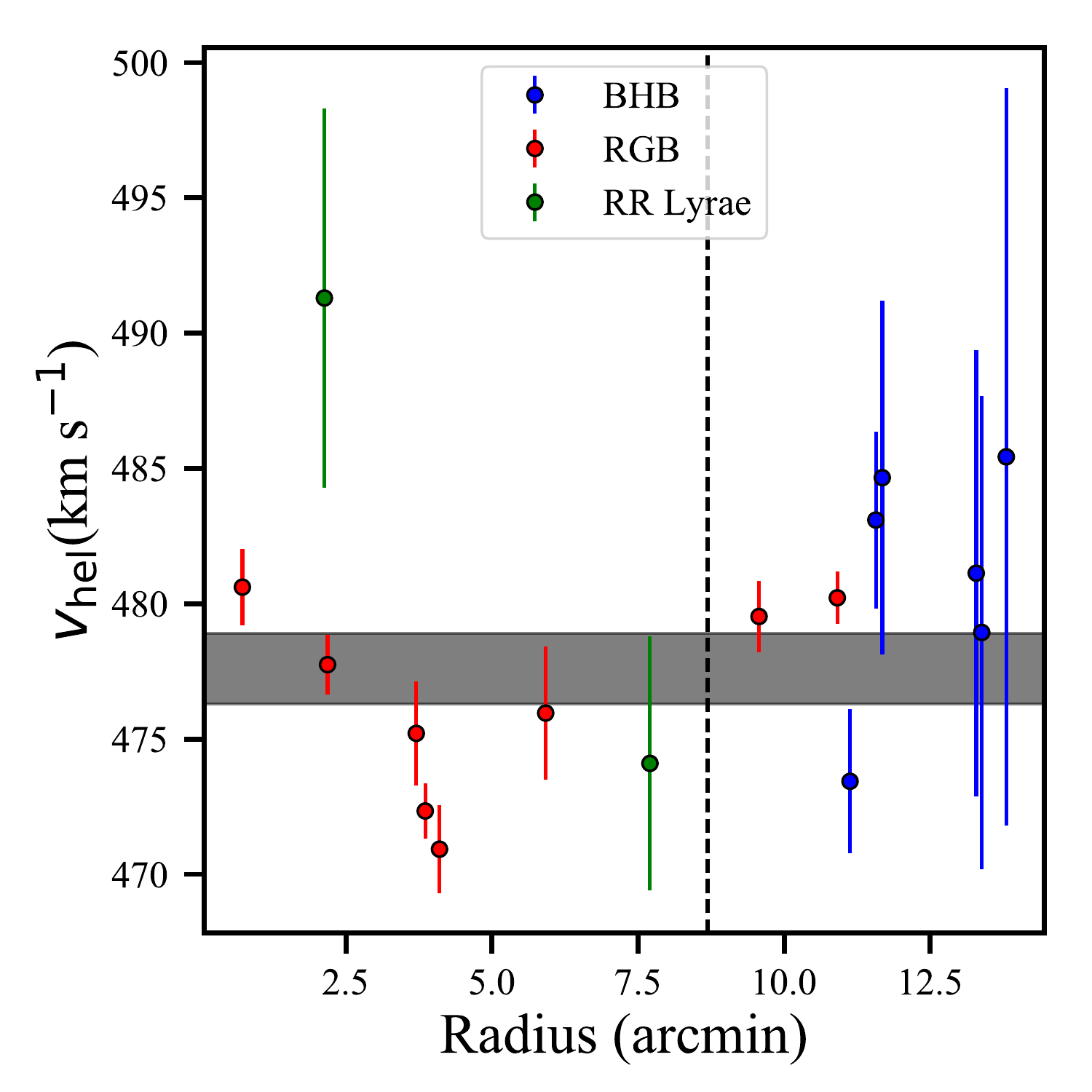}
\caption{Velocity as a function of distance to the center of \CarII, for \CarII members only. Two binary stars are not shown. For the two RR Lyrae stars, the systemic velocities calculated in Sec~\ref{sec:rrl} are used. The six outermost member stars are BHB members at $r > 10\arcmin$. The shaded region shows the systemic velocity of \CarII with $1-\sigma$ uncertainty. The dashed line shows the half-light radius of \CarII from Paper~I. }
\label{Car2_radial}
\end{figure}

We note that the mass estimator from \citet{wolf10} is only valid for dispersion-supported stellar systems in dynamical equilibrium. 
It is possible that \CarII has had a tidal interaction with the Milky Way ($d\sim36$~kpc), LMC ($d\sim20$~kpc), or \CarIII ($d\sim10$~kpc) due to their close proximity. 
Either a velocity gradient or an increasing velocity dispersion at large radii could be potential signs of tidal disruption. 
In \S\ref{sec:vdisp}, we conclude that we are not able to detect a velocity gradient with our current data. 
In Figure~\ref{Car2_radial}, we show the velocity as a function of distance to the center of \CarII.
Interestingly, the six BHB stars are also the outermost members.
As shown in Table~\ref{tab:vdisp_table}, the velocity dispersion from the BHB sample alone is small due to the large velocity uncertainties, and therefore, we also do not see a larger velocity dispersion at large radii.
However, we note that both null results could result from the large velocity uncertainties of these BHB stars. 
Further studies that either identify more RGB stars at large radii or improve the velocity precision for the known BHB members will be necessary to completely rule out a velocity gradient or other tidal effects.

The fact that all 6 BHB members are the outermost stars (Figure~\ref{Car2_radial}) implies a possible non-uniform spatial distribution of the stellar populations in \CarII. 
However, we also caution that this distribution is likely caused by a selection bias from the observations. 
Note that the four outermost BHB members are uniquely identified by AAT, which has the largest FOV among three instruments. 
However, the target selection for AAT was performed with an older version of PARSEC isochrones and therefore may have missed a few RGB members with $r>r_h$, as explained in the Appendix. 
Further observations, including a more comprehensive search for bright RGB members outside of the half-light radius, may be necessary to fully understand the possible stellar population dependent spatial distribution in \CarII.

To check whether the observed kinematics of \CarII and \CarIII are consistent with their being gravitationally bound systems, we have computed their tidal radii ($r_t$) with the Milky Way as the host via equation 18 in \citet{2015MNRAS.453..849B}, using the Milky Way mass model presented in \citet{2016ApJ...829..108E}.  
To set a lower limit on $r_t$, we used the \CarII $M_{\rm 1/2}$ value for the total mass and find  $r_t\sim500$~pc. 
This lower limit on $r_t$ is already significantly larger than the observed size of the system. 
With a mass profile based on a \citet*{nfw96} profile, the total mass of \CarII (with $r = 300$~pc) is estimated to be $5 - 10$ times larger than  $M_{\rm 1/2}$ (with $r_s=0.1-0.5$ kpc), implying $r_t\sim 0.9-1.1$~kpc.
The mass profile of \CarIII is more uncertain due to the small number of stars and the unknown nature of the object. 
If we assume conservatively that \CarIII is a dwarf galaxy with dispersion $\sigma=1\kms$ and $M(r_\mathrm{max}) = 10\times M_{\rm 1/2}$, then we find $r_t\sim250$~pc, again much larger than the observed size of \CarIII.  
If the true dispersion of \CarIII is close to the measurement from a sample of 4 members, then the tidal radius will be even larger.
We additionally computed $r_t$ assuming that the LMC is the host instead of the Milky Way. 
For \CarII, the LMC host $r_t$ values are 5-10\% larger than the Milky Way host values while for \CarIII they are $\sim50\%$ larger.  
Although a complete analysis of the \CarII tidal radius should include the LMC+Milky Way system, we still expect the tidal radius to be larger than the  observed size of \CarII.  
Therefore, we conclude that \CarII is likely to be a bound system based on its current location in the Milky Way, though it is still possible that it had a smaller tidal radius if it approached closer to the Milky Way or LMC in the past.

The small projected separation ($\sim18\arcmin$) of \CarII and \CarIII and their similar distances naturally lead to the question of whether the two are (or were) a bound pair of satellites.
Similar speculation has occurred for the satellite pairs Leo~IV--Leo~V  \citep[$\Delta d_{\rm 3D} \sim 20.6~\kpc$ and $\Delta v \sim 47~\kms$;][]{2008ApJ...686L..83B, 2009ApJ...694L.144W, 2010ApJ...710.1664D} and Pisces~II-Pegasus~III \citep[$\Delta d_{\rm 3D} \sim 43~\kpc$ and $\Delta v \sim 10~\kms$;][]{kim15_peg3, 2016ApJ...833...16K}.
While \CarII and \CarIII have the smallest known physical separation to date, $\Delta d_{\rm 3D} \sim 10~\kpc$, their separation in velocity is quite large $\Delta v \sim 193~\kms$. 

We applied the method presented in \citet{2014MNRAS.440.1225E} to estimate the minimum halo mass for the \CarII-\CarIII system to be bound and find an unrealistically large halo mass of ${\sim}10^{11}~\msun$ (similar to the halo mass of the LMC; \citealt{vanderMarel2014}). 
Based on the observed kinematics of \CarII, the escape velocity at the distance of \CarIII is between $15-25~\kms$, significantly smaller than the observed velocity difference.
\CarII and \CarIII are therefore highly unlikely to be a bound pair of satellites.  

Assuming the pair have similar proper motions, the two satellites would have had a close encounter and sailed past one another ${\sim}53~{\rm Myr}$ ago. 
Based on this trajectory and the observed separation they would have passed within $200$~pc of one another ($\sim2\times(r_{1/2, \, {\rm \CarII}}+r_{1/2, {\rm \CarIII}})$).  
At the point of closest encounter, the \CarIII tidal radius would have been no more than a few tens of parsecs.
Regardless of the nature of \CarIII, a close encounter between the satellites could have disrupted \CarIII. 
While there is no reason to expect \CarII and \CarIII to have similar proper motions given the large difference in their radial velocities, it will be interesting to explore this scenario further when proper motions are available.  We note that the brightest spectroscopic members in both \CarII and \CarIII are brighter than the faint limit for Gaia proper motion measurements.

The properties of \CarII and \CarIII derived in this paper are summarized in Table~\ref{tab:car2_table}.

\subsection{Association with the Magellanic Clouds}
\label{sec:lmcconnect}

As discussed in \S\ref{sec:intro}, the MagLiteS survey was designed to search for satellites of the Magellanic Clouds.
Having searched in the vicinity of the LMC and SMC, it is therefore unsurprising that \CarII and \CarIII are physically close to the Magellanic Clouds. 
The newly measured velocities of the Carina pair can now be used to test whether a physical association with the Clouds is likely.

\begin{figure*}[th!]
\plottwo{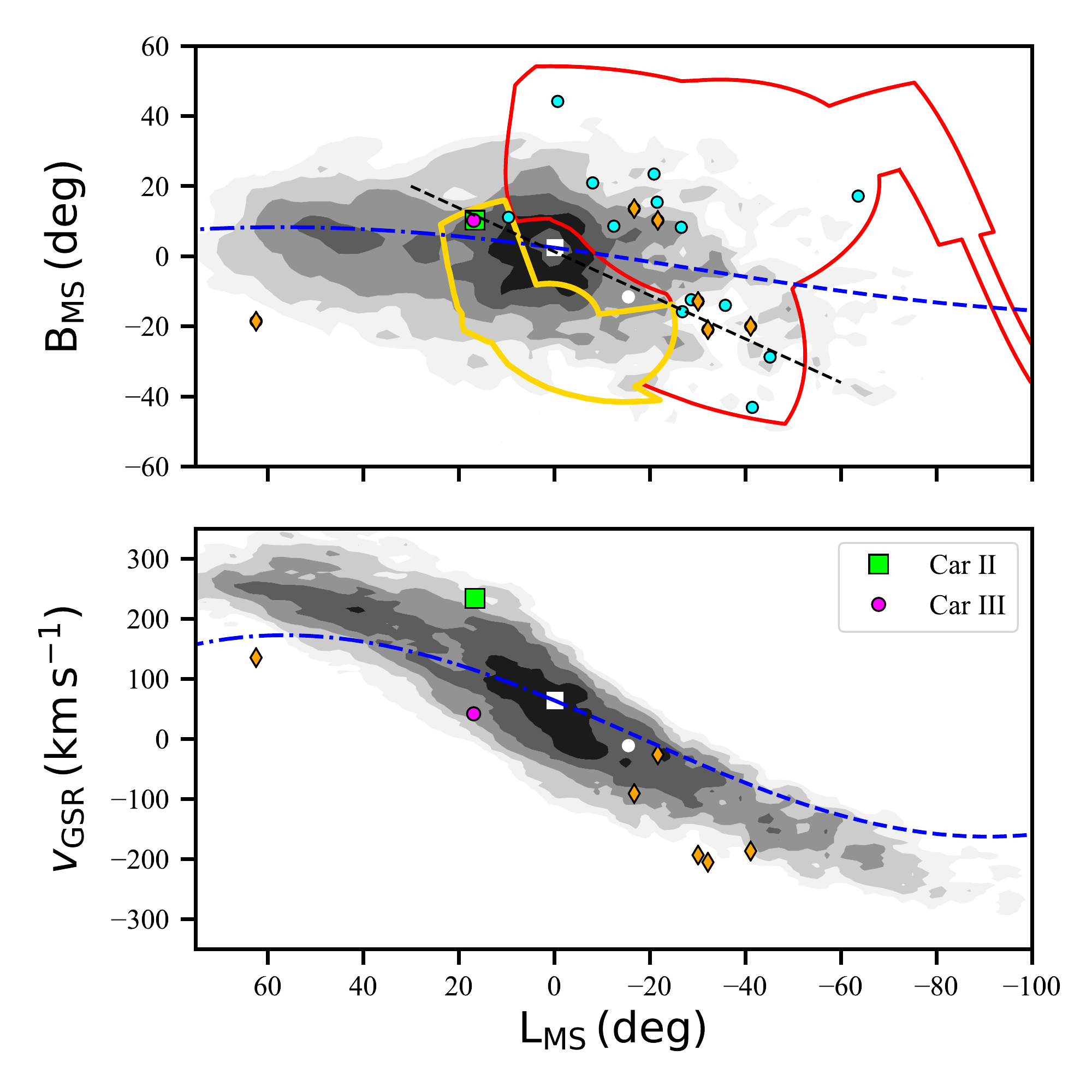}{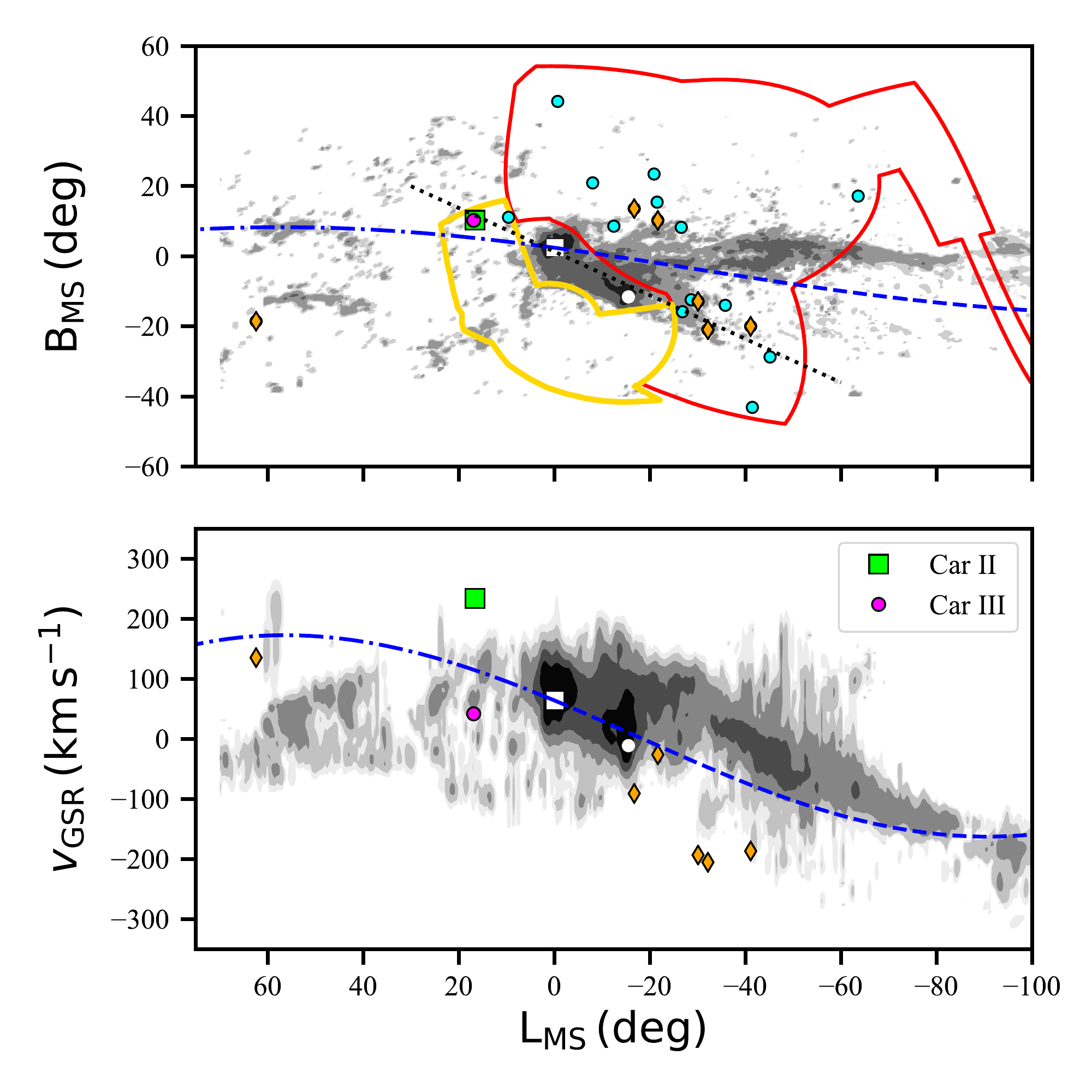}
\caption{\emph{Upper panels:} On-sky positions of the newly discovered satellites in DES (red outline) and MagLiteS (yellow outline), together with the LMC (white square) and SMC (white circle), shown in Magellanic Stream coordinates~\citep{Nidever2008}. \CarII and \CarIII, the Magellanic Clouds, and several other dwarf galaxies form a tight sequence on the sky. The black dashed line is a fit to this sequence as indicated in Paper I.
\emph{Lower panels:} Line-of-sight velocities for ultra-faint satellites near the Magellanic Clouds as a function of Magellanic Stream longitude.
Objects with velocities from the literature are plotted as orange diamonds \citep{kirby15, simon15b, koposov15, 2016ApJ...819...53W, tuc3}, while measurements of \CarII and \CarIII from this work are displayed in green and magenta, respectively. The gray contours show the probability distribution of the LMC satellites from 
\citet{jethwa16} (left) and the neutral hydrogen column density from~\citet{Nidever2010} (right). 
The dash-dotted (dashed) curves show the leading (trailing) orbit of LMC.
}
\label{mc_dist}
\end{figure*}

To aid comparison with models, we first transform the line-of-sight velocities of \CarII and \CarIII from the heliocentric frame to the Galactic Standard of Rest frame\footnote{We adopted the circular orbital velocity of Milky Way at the Sun's radius $\Theta_{0} = 239~\kms$~\citep{McMillan2011} and solar motion of $(U_\odot,~V_\odot,~W_\odot) = (11.1,~12.24,~7.25)~\kms$~\citep{Schonrich2010} for the velocity transformation from heliocentric to Galactic Standard of Rest to match the values used in ~\citet{jethwa16}.} (GSR) and obtain $v_{\rm GSR,~\CarII} = \vgsr$~\kms and $v_{\rm GSR,~\CarIII} = \vgsrb$~\kms. 
Next, we compare these measurements with the dynamical model of Magellanic satellites presented in~\citet{jethwa16}. 
Assuming an association with the LMC, this model predicts a velocity of $v_{\rm GSR} = 118^{+142}_{-80}(149^{+142}_{-114}$~\kms) at the position of \CarII (\CarIII). 
For an association with the SMC, the predicted velocities are higher, at $v_{\rm GSR} = 350^{+50}_{-70}$~\kms for both \CarII and \CarIII. According to this model, both Carinas therefore have velocities consistent with an LMC association. 
In Figure~\ref{mc_dist}, we show the comparison between the observed phase-space distribution of dwarf galaxies/dwarf galaxy candidates and the simulated probability distribution of LMC satellites from the \citet{jethwa16} model, and the neutral hydrogen gas column density from \citet{Nidever2010}. 
According to the \citeauthor{jethwa16} model, both \CarII and \CarIII are consistent with having originated with the LMC.

As pointed out in \S\ref{sec:lmcstar}, the non-rotating LMC halo population and the LMC rotating disk both have $v_\mathrm{hel}\sim380~\kms$ at the location of \CarII and \CarIII, which differs by $\sim$100~\kms from the heliocentric velocities of \CarII and \CarIII.
According to \citet{vanderMarel2014}, the enclosed LMC mass out to a radius of 8.7 kpc is M($<8.7~{\rm kpc)} = 1.7 \times 10^{10} \msun$.  
The corresponding escape velocity is $\sim90~\kms$ at the distance of \CarII ($\sim 18\kpc$ from LMC) and $\sim75~\kms$ at the distance of \CarIII ($\sim 25\kpc$ from LMC). 
Because these values are based on a lower limit to the enclosed mass of the LMC, the actual escape velocities are likely to be somewhat larger. 
Therefore, we tentatively suggest that one or both of \CarII and \CarIII are likely to be bound satellites of the LMC, although proper motion measurements will be needed to confirm this hypothesis.

We also note that all three newly discovered MagLiteS satellite candidates (\CarII, \CarIII, and Pictor II) fall along a linear sequence on the sky as defined by the positions of the LMC, SMC and 7 of the DES satellite candidates. 
This configuration is shown by the dashed black line in Figure~\ref{mc_dist}. 
This linear sequence was first pointed out in~\citet{jethwa16}, prior to any MagLiteS discoveries. 
As discussed in Paper~I, it is unclear whether this linear sequence corresponds to a planar distribution of satellites around the Magellanic Clouds, or simply a satellite distribution that is elongated along the LMC-SMC separation vector. 
Once dynamical models of both scenarios are available, the velocities we have measured may provide a useful discriminant.

\subsection{J and D-factors}
\label{sec:jfactor}

Milky Way satellite galaxies are among the most promising targets for indirect dark matter searches due to their substantial dark matter content, proximity, and dearth of conventional non-thermal emission \citep[e.g.,][]{Baltz:2008wd,Winter2016}.
In particular, analyses of \gammaRayHyph data from the \Fermi Large Area Telescope (LAT) around previously known Milky Way satellites are now sensitive to dark matter annihilating at the canonical thermal relic cross section for particle masses up to 100 \GeV \citep[e.g.,][]{Ackermann2015,GeringerSameth2015}. 
The discovery of additional Milky Way satellites, especially nearby objects such as \CarII and \CarIII, can improve the sensitivity of such searches \citep{He2015,Charles2016}, as demonstrated by \citet{dw15a} and \citet{Albert2017}.

In this subsection, we compute the astrophysical component of the dark matter annihilation and decay fluxes, the so-called J and D-Factors, for both \CarII and \CarIII.  
The J-factor is the line-of-sight integral of the dark matter density squared: $J(\theta) = \int \rho_{\rm DM}^2 \mathrm{d}\Omega \mathrm{d}l$.  
The D-Factor is the linear analog: $D(\theta) = \int \rho_{\rm DM} \mathrm{d}\Omega \mathrm{d}l$. 
Here, $\rho_{\rm DM}$ is the dark matter density and the integral is performed over a solid angle $\Delta \Omega$ with radius $\theta$.
The standard approach for computing $\rho_{\rm DM}$ in dwarf spheroidal galaxies uses the spherical Jeans equation \citep[e.g.,][]{strigari2008, 2015MNRAS.446.3002B}.

The three main ingredients of a spherical Jeans analysis are: the stellar density profile, which we modeled as a Plummer profile \citep{1911MNRAS..71..460P}; the gravitational potential, assumed to be dark matter-dominated and modeled with a Navarro-Frenk-White profile \citep{nfw96}; and the stellar anisotropy, modeled with a constant profile\footnote{Analysis with  generalized stellar, dark matter, and anisotropy profiles would produce larger confidence intervals \citep{2015MNRAS.453..849B}.}. 
Jeffreys priors are assumed for the dark matter halo parameters: $-2 < \log_{10}{\left( r_s/{\rm kpc}\right)} < 1$  and $4 < \log_{10}{\left( \rho_s/{\rm \msun \, kpc^{-3}}\right)} < 14$ for the scale radius, $r_s$, and scale density,   $\rho_s$, respectively.  
Additionally, a prior of $r_s > r_{1/2}$ is imposed, where $r_{1/2}$ is the azimuthally averaged stellar half-light radius.  
We adopted the  $r_s>r_{1/2}$ prior for several reasons: in our posterior distributions there are no trends between $J$ and $r_s$ except for $r_s<r_{1/2}$  where $J$ is systematically higher, the J-factor tends to be overestimated in mock data sets without this cut \citep[see Section 4.1 of][]{2015MNRAS.446.3002B}, and small $r_s$ values are  disfavored in $\Lambda$CDM N-body simulations \citep[based on][ a halo with $V_{\rm max}\sim5-10\kms$ has a $r_s\sim100-300$~pc]{2014MNRAS.444..222G}.
For the anisotropy prior, we assumed a flat symmetrized anisotropy parameter; $-0.95 < \tilde{\beta} < 1.0$ \citep[see Eq. 8 in][]{2006MNRAS.367..387R}.  
A flat prior was used for the average velocity ($470 < \overline{v} < 490 \kms$) and Gaussian priors were assumed for the distance and structural parameters\footnote{We used the azimuthally averaged half-light radius to account for the axisymmetry of the systems ($r_{1/2} = r_{\rm azimuthal} = r_{\rm major}\sqrt{1 -\epsilon}$. For non-spherical analysis of dwarf galaxy J-factors see \citet{2016MNRAS.461.2914H} and \citet{2016PhRvD..94f3521S}.}. 
We used an unbinned likelihood function \citep{2008ApJ...678..614S, Martinez2009JCAP...06..014M, Geringer-Sameth2015ApJ...801...74G} and  determined posterior distributions with the MultiNest sampling routine \citep{2008MNRAS.384..449F, 2009MNRAS.398.1601F}.
We estimated the dark matter $r_t$ (required to compute the J and D-Factors) at each point in the posterior distribution by iteratively computing the enclosed mass and solving for $r_t$ as described in \S\ref{sec:properties}.  
We find the \CarII $r_t$ posterior to be roughly Gaussian, centered at $1~\kpc$ but containing a substantial tail to larger values. 

We calculated the \CarII  integrated J-factor enclosed within solid angles of radii $\theta=\alpha_c, 0.1, 0.2, 0.5^\circ$ to be $\log_{10}{\left(J/{\rm GeV^2 \, cm^{-5}} \right)} = \jcartwoalpha, \jcartwosmall, \jcartwomedium, \jcartwolarge$, respectively, using the 14 star sample.  
$\alpha_c$ is the angle within which the J-factor errors are minimized  \citep{Walker2011ApJ...733L..46W}; $\alpha_c=2 r_{1/2} /d\approx 0.23^\circ$ for \CarII.  
The equivalent radius for the D-Factor occurs at $\alpha_c/2$.
We determined the D-Factor within $\theta=\alpha_c/2, 0.1, 0.2, 0.5^\circ$ to be $\log_{10}{\left(D/{\rm GeV \, cm^{-2}} \right)} = \dcartwoalpha, \dcartwosmall, \dcartwomedium, \dcartwolarge$.
These values agree with the simple J-factor estimator \citep[Eq. 13 of ][]{2016PhRvD..93j3512E}.
This value is an order of magnitude smaller than the simple empirical J-distance scaling relations \citep{dw15a, Albert2017}.
The J-factor contains a large velocity dispersion scaling ($J\propto M^2\propto\sigma^4$) and an increase of only $\Delta \sigma \sim 1.5 \kms$ would move \CarII onto the J-distance scaling relation (Pace \& Strigari in prep).  
There are multiple ultra-faint satellites with larger J-factors \citep[$6-8$ are larger depending on the J-factor compilation;][]{Geringer-Sameth:2014yza, 2015MNRAS.453..849B}.
The D-Factor at $0.1^\circ$ is smaller than most of the other dSphs \citep{2015MNRAS.453..849B}.
Though \CarII has similar distance and velocity dispersion to Ret~II, its J-factor is smaller because it has a larger $r_{1/2}$ (Pace \& Strigari in prep).
\CarII is therefore not the most promising individual target for a dark matter annihilation signal but will be a useful addition in stacked analyses.

We applied the same methodology to the 4 star sample of \CarIII. 
We find the integrated J-factor within solid angles of radii $\theta=\alpha_c, 0.1, 0.2, 0.5^\circ$ to be $\log_{10}{\left(J/{\rm GeV^2 \, cm^{-5}} \right)} = 19.8_{-0.9}^{+1.0}, \jcarthreesmall, \jcarthreemedium, \jcarthreelarge$, respectively. 
$\alpha_c=0.08^\circ$ for \CarIII. 
The D-Factor for \CarIII is $\theta=\alpha_c/2, 0.1, 0.2, 0.5^\circ$ to be $\log_{10}{\left(D/{\rm GeV \, cm^{-2}} \right)} = 17.2_{-0.4}^{+0.5}, \dcarthreesmall, \dcarthreemedium, \dcarthreelarge$.
The J-factor estimation of \CarIII is larger than \CarII due to its proximity, smaller size, and larger (but uncertain) velocity dispersion. 
From our analysis, \CarIII potentially has one of the largest J-factors. 
However, given the very small stellar kinematic sample and the uncertain classification of \CarIII, it is premature to draw strong conclusions about the suitability of \CarIII as a dark matter annihilation target.
As discussed in \S\ref{sec:properties}, the large velocity dispersion could result from binary star motions, small number statistics, or possible tidal effects. 
As a cautionary case in point, the Triangulum~II ultra-faint dwarf galaxy has recently had its J-Factor values revised downward due to the identification of previously unsolved binary stars \citep{2016MNRAS.463.3630G, 2017ApJ...838...83K}. In addition, the velocity dispersion of Bo\"{o}tes~II is likely overestimated in past determinations due to the presence of one binary star~\citep{2009ApJ...690..453K, 2016ApJ...817...41J}.

\begin{figure}[t!]
\epsscale{1.35}
\plotone{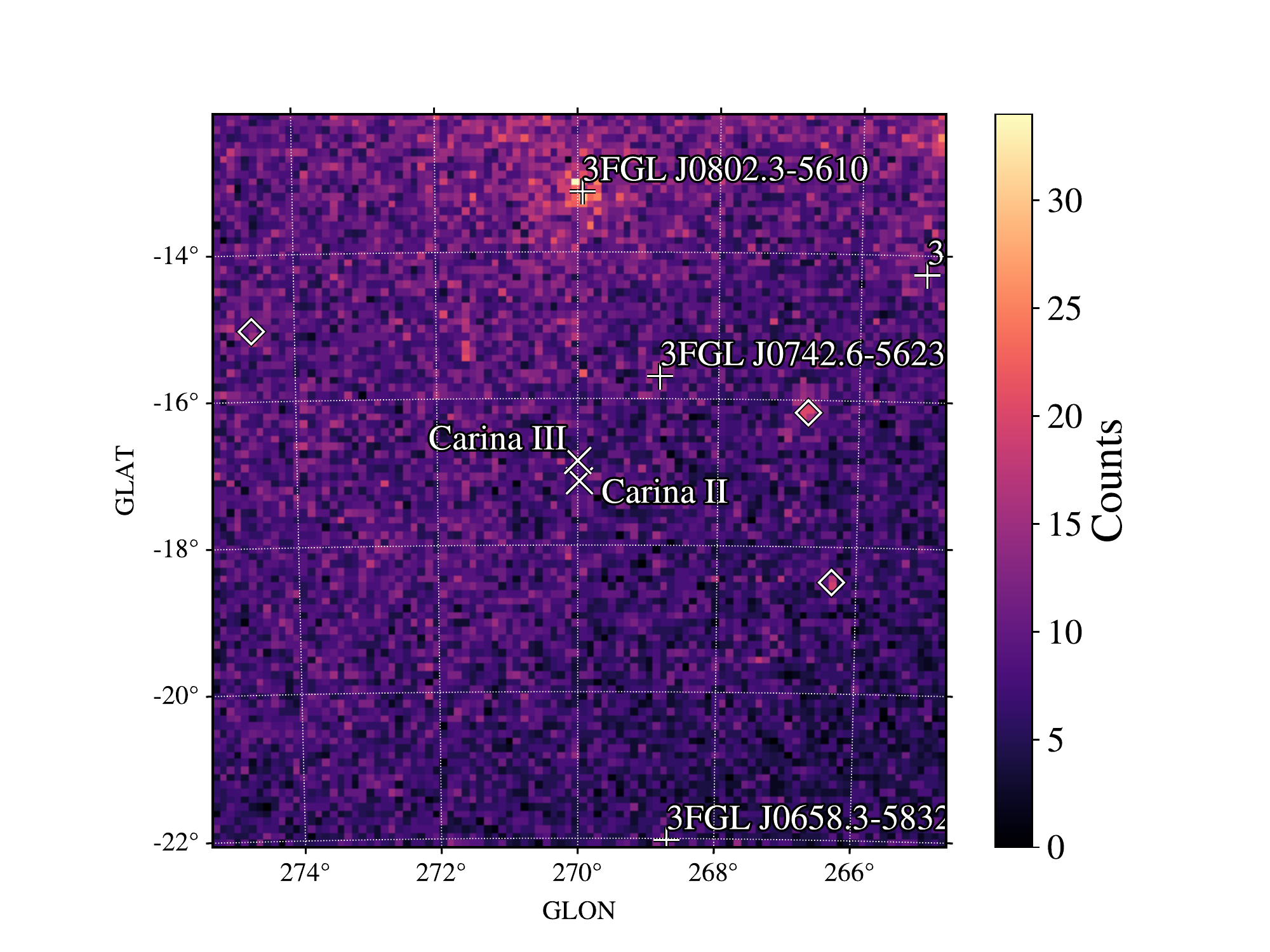}
\caption{
\Fermi LAT \gammaRayHyph counts map ($E > 1 \GeV$) in the vicinity of Car~II and Car~III (Galactic coordinates).  White plus signs indicate the positions of known gamma-ray sources from the 3FGL.  Open diamonds indicate the positions of new gamma-ray point-source candidates found in this analysis.
\label{fermi_counts}
}
\end{figure}

\subsection{Gamma-Ray Observations}
\label{sec:gamma}

We searched for excess \gammaRayHyph emission coincident with \CarII and \CarIII using eight years of LAT data (2008 August 4 to 2016 August 5) passing the \irf{P8R2\_SOURCE}  event class selection from $500\MeV$ to $500\GeV$.
The low-energy bound of 500 \MeV was selected to mitigate the impact of leakage from the bright limb of the Earth because the LAT point spread function broadens considerably below that energy.
The high-energy bound of 500 \GeV is chosen to mitigate the effect of the increasing residual charged-particle background at higher energies~\citep{Ackermann:2014usa}.
To remove \gammaRays produced by cosmic-ray interactions in the Earth's limb, we rejected events with zenith angles greater than $100^{\circ}$.
To analyze data coincident with \CarII and \CarIII, we used $10^{\circ}\times 10^{\circ}$ regions of interest (ROIs) centered on each object.
Data reduction was performed using \stools version 11-05-03.\footnote{\url{http://fermi.gsfc.nasa.gov/ssc/data/analysis/software/}}

We used the maximum-likelihood analysis pipeline described by \citet{Ackermann:2013yva} to test for \gammaRayHyph emission coincident with Car~II and III in excess of the known astrophysical backgrounds.
The background model for the ROI includes Galactic interstellar emission \citep{2016ApJS..223...26A}, isotropic emission\footnote{\url{http://fermi.gsfc.nasa.gov/ssc/data/access/lat/BackgroundModels.html}}, and point sources from a catalog derived from four years of data \citep[3FGL; ][]{3fgl}.
\CarII and \CarIII reside in a region of the sky where the diffuse $\gamma$-ray background is relatively smooth (Galactic latitude of $\sim\,15\degr$), and the nearest 3FGL catalog source is $\sim2\degr$ away.
  
We first created a detection significance map for the entire $10^{\circ}\times 10^{\circ}$ ROI by rastering a putative point source with fixed power-law spectrum ($dN/dE \sim E^{-2}$) across the ROI in $0.1\degr$ steps and computing the improvement in the delta log-likelihood test statistic \citep[TS;][]{Mattox:1996zz}. 
This procedure led to the identification of three additional point-like source candidates in the region, none of which are within $3\degr$ of \CarII or III (Figure~\ref{fermi_counts}).
The TS obtained at the locations of \CarII and \CarIII is 0.16 and 4.2, respectively, both consistent with the background-only hypothesis. 
We note that \CarII and \CarIII would not be resolvable as independent sources given the resolution of the LAT instrument, which is $\sim1\deg$ at 1\GeV and asymptotes to $\sim0.1\deg$ above 10\GeV.
The TSs associated with \CarII and \CarIII are thus correlated, and can be attributed to a single excess of counts located $\sim\,0.7\degr$ from \CarIII  at ($\alpha, \delta$) = (114.5\degr, $-$57.9\degr).

To search for \gammaRayHyph emission consistent with the annihilation of a dark matter particle into standard model products, we fit the \gammaRayHyph data coincident with \CarII and \CarIII using a variety of spectral models generated by DMFit \citep{Jeltema2008,Ackermann:2013yva}.
We scan a range of dark matter masses spanning from $2\GeV$ to $10\TeV$ and annihilating through the \bbbar and \tautau channels.
The most significant excess has TS = 6.2 and occurs for a dark matter particle with mass 35.4 \GeV annihilating through the \tautau channel.
Given that the statistical significance of this excess is well below the typical point-source detection criteria of the LAT (TS = 25), we calculate limits on the dark matter annihilation cross section, \sigmav, using the J-factors derived in \S\ref{sec:jfactor}.
We find that \CarII (\CarIII) can be used to constrain $\sigmav < 2.2 \times 10^{-24} \cm^3 \second^{-1}$ ($3.3 \times 10^{-25} \cm^3 \second^{-1}$) for $100 \GeV$ dark matter particles annihilating through the \bbbar channel. 
These constraints are $\sim 100$ ($\sim 10$) times larger than the thermal-relic cross section \citep[i.e.,][]{Steigman:2012nb}.
We again caution against overinterpreting the constraints derived from \CarIII because the J-factor value of \CarIII is derived from a very small stellar kinematic sample.

\section{SUMMARY}
\label{sec:summary}
In this paper, we presented the first spectroscopic analysis of the Carina~II and Carina~III dwarf galaxy candidates recently discovered in MagLiteS. Based on the kinematic and chemical properties of 18 confirmed spectroscopic member stars in \CarII, we conclude that it is a dark-matter-dominated dwarf galaxy. 
On the other hand, only 4 members were identified in \CarIII. With this small spectroscopic sample we cannot yet determine whether \CarIII is a compact dwarf galaxy or an extended star cluster.

While \CarII and \CarIII are located very close to each other both in sky projection ($\sim18'$) and in three dimensions ($\sim10\kpc$), their systemic velocities differ by $\sim200\kms$. We therefore conclude that these two systems are unlikely to be a pair of bound satellites. 

Both \CarII and \CarIII have line-of-sight velocities consistent with the hypothesis that they formed as members of a group of satellites around the Large Magellanic Cloud (LMC).  
Furthermore, one or both systems might remain bound to LMC due to the small difference in the line-of-sight velocity.
The brightest RGB members in \CarII and \CarIII are bright enough to have proper motion measurements from Gaia to test this hypothesis.

We further identify one blue horizontal branch (BHB) star as a likely LMC member in the region of \CarII. Located at about 18\degr\ from the center of the LMC, this star is one of the LMC's most distant spectroscopically confirmed BHB members, and might provide hint on our understanding of the structure and dynamics of the LMC's outer regions.

No statistically significant excess of \gammaRayHyph emission is found at the locations of \CarII and \CarIII in eight years of \textit{Fermi}-LAT data. Using the J-factors derived from the kinematics data,  \CarII and \CarIII can be used to constrain the dark matter annihilation cross section.

\acknowledgements{
TSL thanks Leo Girardi for helpful discussions regarding the PARSEC isochrones. The authors thank Louis Strigari for helpful conversations regarding the J-factor calculation.
We are grateful for the service observations and Director's Discretionary time on AAT and VLT.
We thank the service observers, Jeffrey Simpson and Chris Lidman, for collecting the AAT data during the service runs.  
We acknowledge the traditional owners of the land on which the AAT stands, the Gamilaraay people, and pay our respects to elders past and present. 

This project is partially supported by the NASA Fermi Guest Investigator Program Cycle 9 No. 91201. 
JDS acknowledges support from the National Science Foundation under grant AST-1714873. 
ABP acknowledges generous support from the George P. and Cynthia Woods Institute for Fundamental Physics and Astronomy at Texas A\&M University. 
MASC is supported by the {\it Atracci\'on de Talento} contract no. 2016-T1/TIC-1542 granted by the Comunidad de Madrid in Spain.
APJ is supported by NASA through Hubble Fellowship grant HST-HF2-51393.001 awarded by the Space Telescope Science Institute, which is operated by the Association of Universities for Research in Astronomy, Inc., for NASA, under contract NAS5-26555. 
BCC acknowledges the support of the Australian Research Council through Discovery project DP150100862.

This research has made use of NASA's Astrophysics Data System Bibliographic Services.
This research made use of \code{Astropy}, a community-developed core Python package for Astronomy~\citep{Astropy2013}.
Contour plots were generated using \code{corner.py} \citep{corner}.

This project used data obtained with the Dark Energy Camera (DECam),
which was constructed by the Dark Energy Survey (DES) collaboration.
Funding for the DES Projects has been provided by 
the U.S. Department of Energy, 
the U.S. National Science Foundation, 
the Ministry of Science and Education of Spain, 
the Science and Technology Facilities Council of the United Kingdom, 
the Higher Education Funding Council for England, 
the National Center for Supercomputing Applications at the University of Illinois at Urbana-Champaign, 
the Kavli Institute of Cosmological Physics at the University of Chicago, 
the Center for Cosmology and Astro-Particle Physics at the Ohio State University, 
the Mitchell Institute for Fundamental Physics and Astronomy at Texas A\&M University, 
Financiadora de Estudos e Projetos, Funda{\c c}{\~a}o Carlos Chagas Filho de Amparo {\`a} Pesquisa do Estado do Rio de Janeiro, 
Conselho Nacional de Desenvolvimento Cient{\'i}fico e Tecnol{\'o}gico and the Minist{\'e}rio da Ci{\^e}ncia, Tecnologia e Inovac{\~a}o, 
the Deutsche Forschungsgemeinschaft, 
and the Collaborating Institutions in the Dark Energy Survey. 
The Collaborating Institutions are 
Argonne National Laboratory, 
the University of California at Santa Cruz, 
the University of Cambridge, 
Centro de Investigaciones En{\'e}rgeticas, Medioambientales y Tecnol{\'o}gicas-Madrid, 
the University of Chicago, 
University College London, 
the DES-Brazil Consortium, 
the University of Edinburgh, 
the Eidgen{\"o}ssische Technische Hoch\-schule (ETH) Z{\"u}rich, 
Fermi National Accelerator Laboratory, 
the University of Illinois at Urbana-Champaign, 
the Institut de Ci{\`e}ncies de l'Espai (IEEC/CSIC), 
the Institut de F{\'i}sica d'Altes Energies, 
Lawrence Berkeley National Laboratory, 
the Ludwig-Maximilians Universit{\"a}t M{\"u}nchen and the associated Excellence Cluster Universe, 
the University of Michigan, 
{the} National Optical Astronomy Observatory, 
the University of Nottingham, 
the Ohio State University, 
the OzDES Membership Consortium
the University of Pennsylvania, 
the University of Portsmouth, 
SLAC National Accelerator Laboratory, 
Stanford University, 
the University of Sussex, 
and Texas A\&M University.

Based on observations at Cerro Tololo Inter-American Observatory, National Optical
Astronomy Observatory (NOAO Prop. ID 2016A-0366 and PI Keith Bechtol), which is operated by the Association of
Universities for Research in Astronomy (AURA) under a cooperative agreement with the
National Science Foundation.

}

{\it Facilities:} 
 \facility{Magellan/Baade (IMACS); Very Large Telescope (GIRAFEE+FLAME); Anglo-Australian Telescope (AAOmega+2dF)}

\bibliographystyle{apj}
\bibliography{main}{}

\clearpage
\appendix
\section{PARSEC isochrones}
\label{iso}

In the course of our spectroscopic follow-up campaign on DES ultra-faint satellites \citep{simon15b, tuc3, eri2}, we noticed that the location of the confirmed spectroscopic members in color-magnitude diagrams was shifted with respect to the PARSEC isochrones of a metal-poor stellar population in the DECam system ~\citep[see e.g., Figure~1 in][]{tuc3}. 
In particular, the confirmed RGB members are found to have bluer colors than indicated by PARSEC isochrones. This offset is on the order of $g-r\sim0.05$~mag and could significantly affect the target selection of the candidate members for spectroscopic follow-up. For the IMACS observations presented in this paper, the target selection used criteria similar to those applied by \citet{tuc3} and therefore the effect caused by this shift was minimal. However, for AAT observations, targets were selected to be near the PARSEC isochrone and therefore most of the RGB members were missed as a result of this color offset. We therefore manually shifted the isochrone bluer for the target selection of VLT observation, as described in \S\ref{sec:giraffe}. 

\begin{figure}[th!]
\plottwo{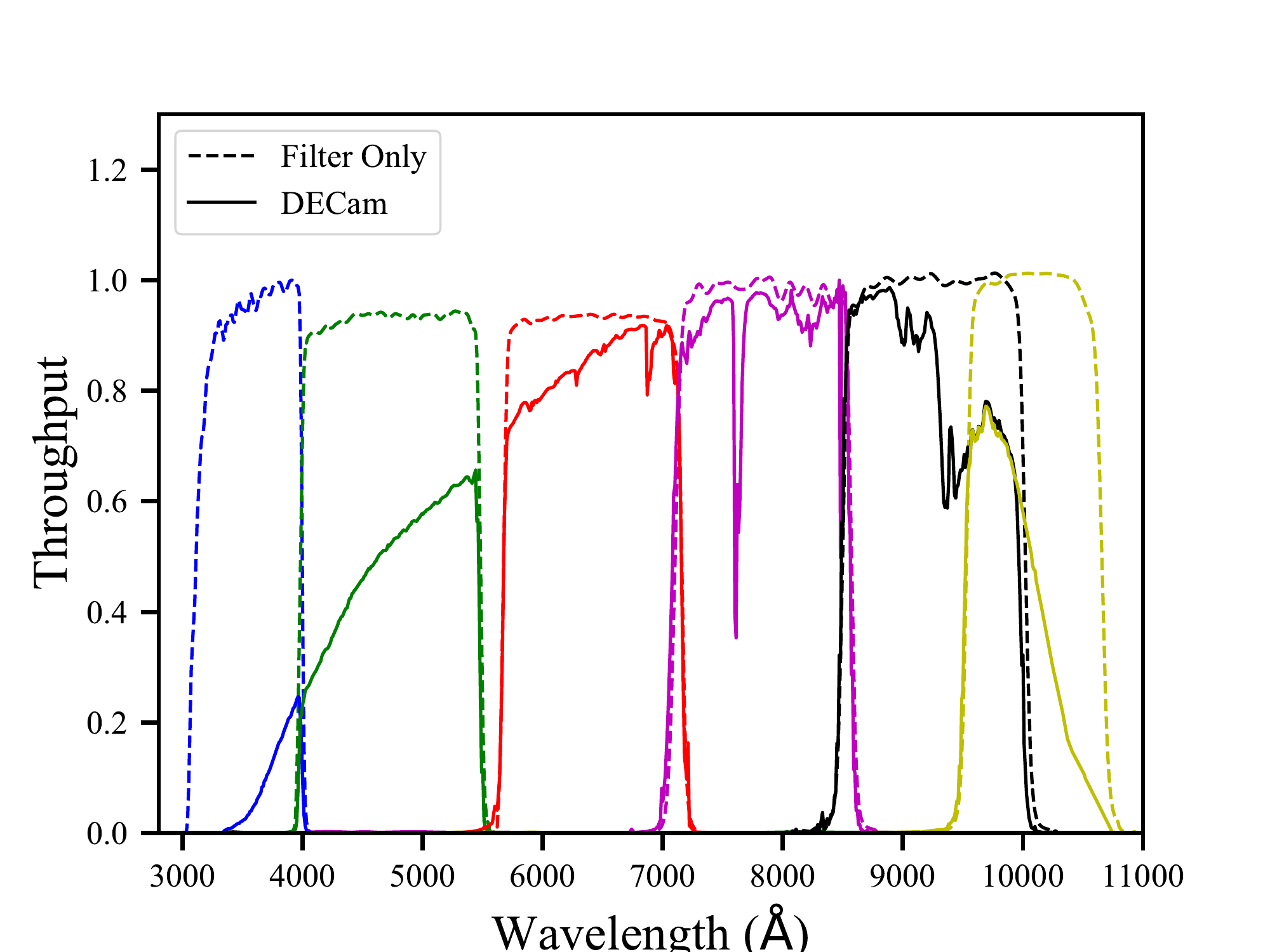}{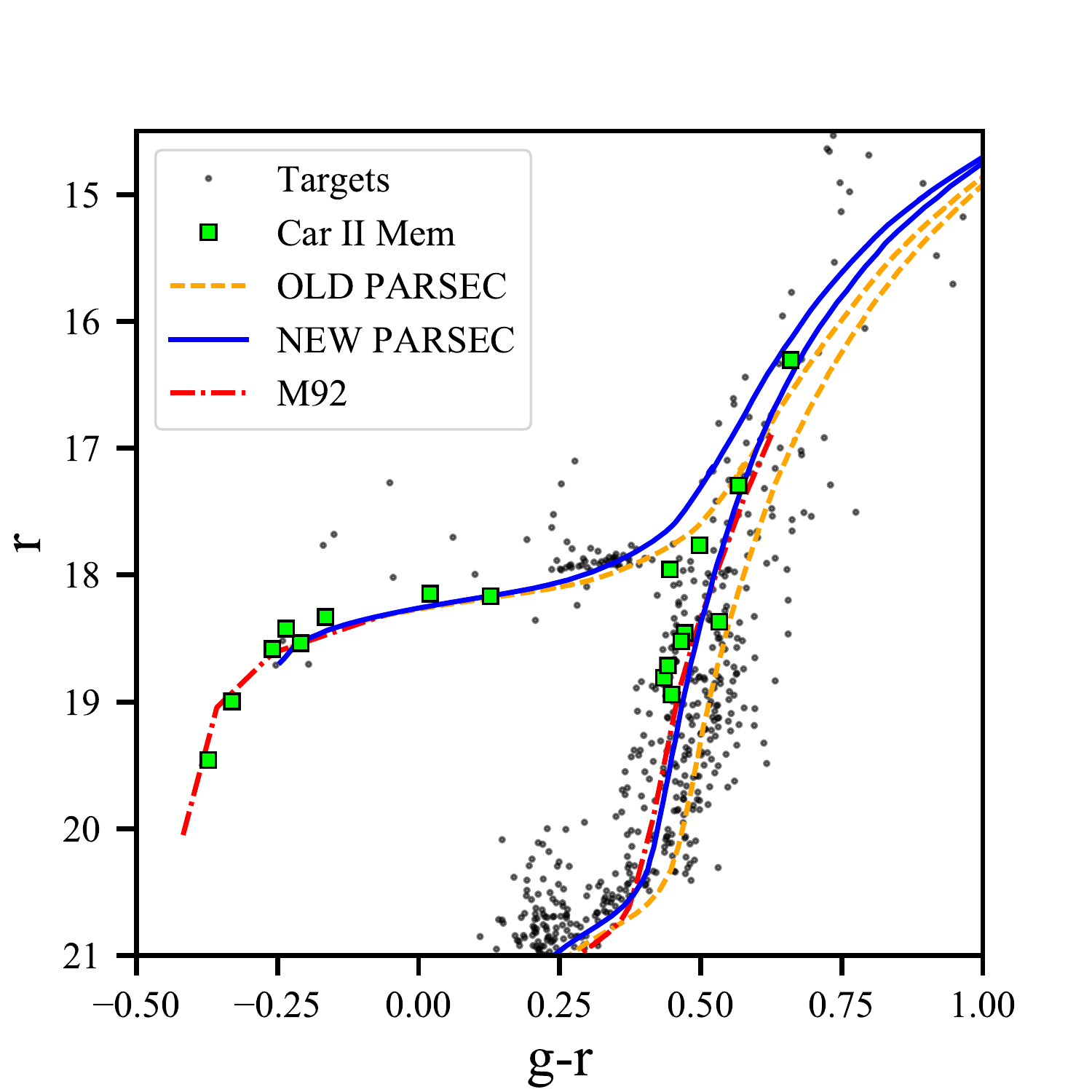}
\caption{($left$) A comparison of the DECam system throughput including filter response only (old) and combined (filter+CCD QE+atmosphere, new), both from CTIO website. ($right$) PARSEC isochrones of a metal-poor population with \feh = $-$2.2, age = 12 Gyr using the old and new DECam response. The new isochrone is well aligned with the confirmed members in \CarII as well as the empirical isochrone of M92 from~\citet{an08} translated from SDSS to the DECam system (red dashed lines).}
\label{decam}
\end{figure}

This offset was only apparent when comparing confirmed members of ultra-faint dwarf galaxies discovered with DECam to the PARSEC isochrone generated in the DECam system. A similar offset is not observed for the ultra-faint dwarfs discovered in SDSS. After consultation with the Padova team, we learned that the PARSEC isochrones were computed using the filter transmission of DECam only. However, in order to compute precise synthetic magnitudes, one should also consider the CCD quantum efficiency (QE), Earth's atmospheric transmission, and the mirror reflectivity, etc. A newer version of the PARSEC models for the DECam system was updated\footnote{PARSEC isochrones at \url{http://stev.oapd.inaf.it/cgi-bin/cmd}.} on 2017 July 17 with the latest version of the system response of DECam\footnote{Details of the DECam system response and filter throughput are available at CTIO website \url{http://www.ctio.noao.edu/noao/node/1033}. The specific system response file used to compute the new isochrones is available at \url{http://www.ctio.noao.edu/noao/sites/default/files/DECam/DECam\_filters.xlsx}.}. A comparison of different versions of DECam throughput and the corresponding PARSEC isochrones are shown in Figure~\ref{decam}. With the corrected DECam system response, the confirmed spectroscopic members in \CarII are much closer to the updated PARSEC isochrone. The remaining small offset might be caused by an overestimate of the interstellar extinction\footnote{All photometry presented in this paper is corrected for interstellar extinction using the extinction map from~\citet{1998ApJ...500..525S}. The E(B-V) value around \CarII and \CarIII is around 0.19}.

\clearpage
\clearpage
\begin{turnpage}
\begin{deluxetable*}{c c c c c c c c l c c c}
\tabletypesize{\tiny}
\tablecaption{Velocity and metallicity measurements.
\label{tab:car2_spec}
}
\tablehead{ID & $\alpha_{2000}$ & $\delta_{2000}$  & $g_0$\tablenotemark{a} & $r_0$\tablenotemark{a} & Masks/ & MJD &  S/N & \multicolumn{1}{c}{$v$} & ${\rm EW}$ & ${\rm [Fe/H]}$ & Comment\\ 
 & (deg) & (deg) & (mag) & (mag) & Instruments  &  &  & \multicolumn{1}{c}{(\kms)} & (\AA) &  & }
\startdata
 \multicolumn{12}{c}{ \ruleline{Carina II} } \\
MAGLITES\,J073504.92$-$575646.9 & 113.77049 & $-$57.94636 & 18.18 & 18.35 & \code{AAT-Jan} & 57777.7 &  7.9 & $482.80 \pm 3.38$ & \noinfo & \noinfo &        BHB\\
    &     &     &     &     & \code{AAT-May} & 57902.4 &  3.3 & $487.20 \pm 12.63$ & \noinfo & \noinfo &    \\
MAGLITES\,J073546.15$-$574911.6 & 113.94231 & $-$57.81988 & 18.68 & 19.02 & \code{AAT-Jan} & 57777.7 &  4.4 & $484.66 \pm 6.53$ & \noinfo & \noinfo &        BHB\\
MAGLITES\,J073558.28$-$580328.2 & 113.99285 & $-$58.05784 & 19.25 & 18.81 & \code{IMACS-Car2Mask2} & 57778.3 &  4.8 & $475.96 \pm 2.47$ & $1.78 \pm 0.28$ & $-2.74 \pm 0.18$ &    \\
MAGLITES\,J073558.39$-$581227.7 & 113.99329 & $-$58.20769 & 18.36 & 18.61 & \code{AAT-Jan} & 57777.7 &  4.1 & $485.43 \pm 13.64$ & \noinfo & \noinfo &        BHB\\
MAGLITES\,J073559.15$-$575918.2 & 113.99644 & $-$57.98838 & 18.93 & 18.45 & \code{IMACS-Car2Mask2} & 57778.3 &  7.3 & $470.93 \pm 1.62$ & $2.45 \pm 0.24$ & $-2.43 \pm 0.13$ &    \\
MAGLITES\,J073601.33$-$575843.8 & 114.00552 & $-$57.97885 & 18.26 & 17.76 & \code{VLT-Feb} & 57811.1 & 32.9 & $472.34 \pm 1.03$ & $2.99 \pm 0.28$ & $-2.32 \pm 0.14$ &    \\
MAGLITES\,J073611.87$-$581228.6 & 114.04945 & $-$58.20796 & 18.23 & 18.46 & \code{AAT-Jan} & 57777.7 &  6.0 & $478.94 \pm 8.75$ & \noinfo & \noinfo &        BHB\\
MAGLITES\,J073621.25$-$575800.3 & 114.08852 & $-$57.96675 & 16.98 & 16.30 & \code{AAT-Jan} & 57777.7 & 40.7 & $492.70 \pm 0.51$ & $4.30 \pm 0.26$ & $-2.07 \pm 0.12$ &     Binary\\
    &     &     &     &     & \code{IMACS-Car2Mask2} & 57778.3 & 42.4 & $494.30 \pm 1.03$ & $4.66 \pm 0.22$ & $-1.92 \pm 0.10$ &    \\
    &     &     &     &     & \code{AAT-May} & 57902.4 & 20.1 & $464.92 \pm 0.83$ & $4.44 \pm 0.32$ & $-2.01 \pm 0.14$ &    \\
MAGLITES\,J073624.62$-$575922.1 & 114.10256 & $-$57.98948 & 18.89 & 18.36 & \code{IMACS-Car2Mask2} & 57778.3 &  7.3 & $479.46 \pm 2.10$ & \noinfo & \noinfo &    \\
    &     &     &     &     & \code{IMACS-Car2Mask1} & 57779.3 &  8.3 & $481.56 \pm 1.90$ & $2.24 \pm 0.90$ & $-2.56 \pm 0.48$ &    \\
MAGLITES\,J073624.98$-$575714.3 & 114.10408 & $-$57.95397 & 18.43 & 17.97 & \code{VLT-Feb} & 57811.1 & 31.5 & $477.75 \pm 1.11$ & $2.01 \pm 0.20$ & $-2.77 \pm 0.12$ &    \\
MAGLITES\,J073637.00$-$580114.5 & 114.15416 & $-$58.02069 & 18.56 & 18.36 & \code{AAT-Jan} & 57777.7 &  9.0 & $489.30 \pm 1.84$ & \noinfo & \noinfo &   RR Lyrae\\
    &     &     &     &     & \code{IMACS-Car2Mask2} & 57778.3 &  9.3 & $457.03 \pm 2.25$ & \noinfo & \noinfo &    \\
    &     &     &     &     & \code{IMACS-Car2Mask1} & 57779.3 &  8.1 & $498.71 \pm 1.63$ & \noinfo & \noinfo &    \\
    &     &     &     &     & \code{AAT-May} & 57902.4 &  4.6 & $440.82 \pm 14.58$ & \noinfo & \noinfo &    \\
MAGLITES\,J073645.86$-$575154.1 & 114.19109 & $-$57.86503 & 17.97 & 18.02 & \code{AAT-Jan} & 57777.7 &  8.3 & $484.83 \pm 2.01$ & \noinfo & \noinfo &   RR Lyrae\\
    &     &     &     &     & \code{IMACS-Car2Mask1} & 57779.3 & 10.4 & $457.63 \pm 2.06$ & \noinfo & \noinfo &    \\
MAGLITES\,J073646.47$-$575910.2 & 114.19362 & $-$57.98617 & 19.39 & 18.93 & \code{AAT-Jan} & 57777.7 &  3.6 & $485.57 \pm 4.42$ & \noinfo & \noinfo &     Binary\\
    &     &     &     &     & \code{IMACS-Car2Mask2} & 57778.3 &  5.4 & $484.15 \pm 3.17$ & \noinfo & \noinfo &    \\
    &     &     &     &     & \code{IMACS-Car2Mask1} & 57779.3 &  5.0 & $486.62 \pm 2.59$ & $2.62 \pm 0.29$ & $-2.26 \pm 0.15$ &    \\
    &     &     &     &     & \code{VLT-Feb} & 57811.1 & 17.5 & $460.69 \pm 1.37$ & $2.82 \pm 0.30$ & $-2.16 \pm 0.16$ &    \\
MAGLITES\,J073655.60$-$580049.8 & 114.23168 & $-$58.01383 & 18.99 & 18.52 & \code{IMACS-Car2Mask2} & 57778.3 &  7.2 & $475.21 \pm 1.92$ & \noinfo & \noinfo &    \\
MAGLITES\,J073729.30$-$580447.8 & 114.37206 & $-$58.07993 & 19.14 & 18.68 & \code{VLT-Feb} & 57811.1 & 19.6 & $479.53 \pm 1.32$ & $2.42 \pm 0.33$ & $-2.40 \pm 0.18$ &    \\
MAGLITES\,J073737.04$-$574925.5 & 114.40434 & $-$57.82375 & 19.09 & 19.46 & \code{VLT-Feb} & 57811.1 &  6.0 & $481.13 \pm 8.25$ & \noinfo & \noinfo &        BHB\\
MAGLITES\,J073739.81$-$580507.0 & 114.41589 & $-$58.08528 & 17.80 & 17.25 & \code{VLT-Feb} & 57811.1 & 46.8 & $480.22 \pm 0.97$ & $2.52 \pm 0.26$ & $-2.64 \pm 0.13$ &    \\
MAGLITES\,J073745.86$-$580406.7 & 114.44108 & $-$58.06853 & 18.28 & 18.52 & \code{AAT-Jan} & 57777.7 &  4.1 & $467.44 \pm 8.43$ & \noinfo & \noinfo &        BHB\\
    &     &     &     &     & \code{VLT-Feb} & 57811.1 & 16.9 & $474.10 \pm 2.80$ & \noinfo & \noinfo &    \\
    \multicolumn{12}{c}{ \ruleline{Carina III} } \\
MAGLITES\,J073823.68$-$575150.8 & 114.59866 & $-$57.86412 & 20.21 & 19.71 & \code{VLT-Feb} & 57811.1 &  6.4 & $290.56 \pm 4.99$ & \noinfo & \noinfo &     \\
MAGLITES\,J073834.84$-$575211.2 & 114.64516 & $-$57.86977 & 17.60 & 17.77 & \code{AAT-Jan} & 57777.7 &  8.5 & $277.14 \pm 3.93$ & \noinfo & \noinfo &        BHB\\
    &     &     &     &     & \code{IMACS-Car3LongSlit} & 57779.3 &  6.9 & $282.26 \pm 2.99$ & \noinfo & \noinfo &    \\
MAGLITES\,J073834.94$-$575705.4 & 114.64558 & $-$57.9515 & 17.70 & 17.18 & \code{IMACS-Car3Mask1} & 57778.2 & 29.9 & $280.19 \pm 1.29$ & $3.73 \pm 0.25$ & $-1.97 \pm 0.12$ &     \\
    &     &     &     &     & \code{VLT-Feb} & 57811.1 & 43.8 & $280.36 \pm 1.00$ & $3.73 \pm 0.28$ & $-1.97 \pm 0.13$ &    \\
MAGLITES\,J073835.54$-$575622.3 & 114.64808 & $-$57.93952 & 17.53 & 17.68 & \code{IMACS-Car3Mask1} & 57778.2 & 15.6 & $288.25 \pm 1.56$ & \noinfo & \noinfo &        BHB\\
    &     &     &     &     & \code{VLT-Feb} & 57811.1 & 23.8 & $291.92 \pm 1.86$ & \noinfo & \noinfo &    \\
    &     &     &     &     & \code{AAT-May} & 57902.4 &  4.4 & $288.25 \pm 4.92$ & \noinfo & \noinfo &    \\
\multicolumn{12}{c}{ \ruleline{Non Member\tablenotemark{b}} } \\
MAGLITES\,J073507.41$-$574725.4 & 113.78087 & $-$57.79038 & 18.13 & 17.89 & \code{AAT-Jan} & 57777.7 & 13.5 & $291.13 \pm 1.80$ & $3.56 \pm 0.56$ & \noinfo &     \\
    &     &     &     &     & \code{AAT-May} & 57902.4 &  7.3 & $298.29 \pm 2.30$ & \noinfo & \noinfo &    \\
MAGLITES\,J073613.95$-$580641.2 & 114.05814 & $-$58.11145 & 18.31 & 17.92 & \code{AAT-Jan} & 57777.7 & 10.0 & $396.63 \pm 1.81$ & $2.64 \pm 0.63$ & \noinfo &     \\
    &     &     &     &     & \code{AAT-May} & 57902.4 &  6.9 & $388.58 \pm 2.72$ & \noinfo & \noinfo &    \\
MAGLITES\,J073629.58$-$574940.0 & 114.12327 & $-$57.82777 & 18.21 & 17.96 & \code{AAT-Jan} & 57777.7 &  8.9 & $309.01 \pm 2.19$ & $3.55 \pm 0.36$ & \noinfo &     \\
MAGLITES\,J073634.86$-$580340.6 & 114.14526 & $-$58.06127 & 18.51 & 18.70 & \code{AAT-Jan} & 57777.7 &  5.5 & $329.64 \pm 3.85$ & \noinfo & \noinfo &        BHB\\
    &     &     &     &     & \code{IMACS-Car2Mask2} & 57778.3 &  4.0 & $327.09 \pm 3.91$ & \noinfo & \noinfo &    \\
    &     &     &     &     & \code{VLT-Feb} & 57811.1 & 14.4 & $335.29 \pm 2.87$ & \noinfo & \noinfo &    \\
MAGLITES\,J073651.54$-$580247.8 & 114.21475 & $-$58.04662 & 18.07 & 17.79 & \code{IMACS-Car2Mask2} & 57778.3 & 11.8 & $294.38 \pm 1.41$ & $4.29 \pm 0.39$ & \noinfo &     \\
MAGLITES\,J073710.30$-$575819.3 & 114.29293 & $-$57.97202 & 19.92 & 19.55 & \code{VLT-Feb} & 57811.1 &  9.5 & $261.93 \pm 3.18$ & $4.20 \pm 0.59$ & \noinfo &     \\
MAGLITES\,J073727.87$-$574421.9 & 114.36613 & $-$57.73943 & 18.21 & 17.93 & \code{AAT-Jan} & 57777.7 & 11.3 & $266.26 \pm 1.66$ & $5.17 \pm 0.44$ & \noinfo &     \\
    &     &     &     &     & \code{AAT-May} & 57902.4 &  5.2 & $294.10 \pm 4.91$ & \noinfo & \noinfo &    \\
    \enddata
\tablecomments{ (a) Quoted magnitudes represent the weighted-average PSF magnitude derived from the original MagLiteS survey (rather than the photometry from time-series follow-up observations that were used to search for RR Lyrae stars). The photometry provided here is the dereddened photometry using the extinction map from~\citet{1998ApJ...500..525S}. A the location of \CarII and \CarIII, the average color excess is $E(B-V)\sim0.19$. \\
(b) For non-members, only stars with $v_\mathrm{hel} > 220~\kms$ are presented here. The remaining non-members are available in the online version in machine readable format.
}
\end{deluxetable*}

\end{turnpage}

\end{document}